\documentclass[twocolumn,floatfix,amsfonts,amssymb,notitlepage]{revtex4-1}
\usepackage{color}
\usepackage{graphicx}
\usepackage{amsmath}
\usepackage{latexsym}
\usepackage{epstopdf}
\usepackage{amsmath}
\usepackage{amssymb}
\usepackage{dsfont}
\usepackage{multirow}
\usepackage{mathtools}
\usepackage{tcolorbox}
\usepackage{ulem}

\newcommand{\beq}{\begin{equation}}
\newcommand{\eeq}{\end{equation}}
\newcommand{\bea}{\begin{eqnarray}}
\newcommand{\eea}{\end{eqnarray}}

\def\bra#1{\langle #1|}
\def\ket#1{|#1\rangle}

\newcommand{\bebox}{\begin{tcolorbox}}
\newcommand{\eebox}{\end{tcolorbox}}

\newcommand{\eq}{\begin{equation}}
\newcommand{\en}{\end{equation}}
\newcommand{\ear}{\begin{eqnarray}}
\newcommand{\rae}{\end{eqnarray}}

\newcommand{\rf}[1]{(\ref{#1})}
\newcommand {\inpar}[1]{\lfloor{#1}\rfloor}
\newcommand {\mbar}{\overline{M}}
\newcommand{\scp}{\scriptsize}
\def\bra#1{\langle #1|}
\def\ket#1{|#1\rangle}

\newcommand{\id}{\mathds{1}}
\newcommand{\be}{\begin{eqnarray}}
\newcommand{\ee}{\end{eqnarray}}

\makeindex

\begin{document}
\title{Conformally invariant free parafermionic quantum chains with multispin interactions}
\author{Francisco C. Alcaraz Lucas M. Ramos}
\email{alcaraz@ifsc.usp.br}
\email{lucas.morais@ifsc.usp.br}
\affiliation{ Instituto de F\'{\i}sica de S\~{a}o Carlos, Universidade de S\~{a}o Paulo,
Caixa Postal 369, 13560-970, S\~{a}o Carlos, SP, Brazil}
%\author{Lucas M. Ramos}
%\affiliation{Departamento de F\'isica, 
%          Universidade Federal de Lavras, 
%          Caixa Postal 3037, 37200-000, Lavras, MG, Brazil}
\date{\today{}}

\begin{abstract}

  We calculated the spectral properties of two related families of
non-Hermitian free-particle quantum chains with $N$-multispin
interactions ($N=2,3,\ldots$). The first family have a $Z(N)$
symmetry and are described by free parafermions. The second one
has a $U(1)$ symmetry and are generalizations of $XX$ quantum chains
described by free fermions. The eigenspectra of both free-particle
families are formed by the combination of the same pseudo-energies.
The models have a multicritical point with dynamical critical
exponent $z=1$. The finite-size behavior of their eigenspectra,
as well as the entanglement properties of their ground state
wave function, indicate the models are conformally invariant.
The models with open and periodic boundary conditions show
quite distinct physics due to their non-Hermiticity. The models
defined with open boundaries have a single conformal invariant
phase while the $XX$ multispin models show multiple phases with 
distinct conformal central charges in the
periodic case. The critical exponents of the models are calculated
for $N=3,4,5$ and $6$.
~                        

\end{abstract}

\maketitle

\section{Introduction}

Exactly integrable quantum chains with a free-particle eigenspectrum plays an 
important role in the understanding of many body physics. They are simple models 
and in general are solved by the standard Jordan-Wigner transformation \cite{lieb,pfeuty}. Thanks 
to this transformation most of these models are mapped into an effective Hamiltonian 
formed by the addition of bilinear fermionic operators, whose solutions follows 
 from a generalized Fourier transformation (Bogoliubov transformation). Recently a 
large class of  
 free-particle quantum chains, that are not bilinear after the Jordan-Wigner 
transformation were introduced. These are models defined in terms of $Z(N)$  
parafermionic operators ($N=2,3,\ldots$), containing multispin interactions 
involving ($p+1$)-spins  ($p=1,2,\ldots$). The exact solutions  are known 
only when the quantum chains are defined in a lattice with open boundary condition 
(OBC). In the case $p=1$ and $N=2$ they recover the standard free-fermionic quantum chain with two spin interactions. The simplest case, where $p=2$ and $N=2$,
 is the 
free fermionic Fendley 3-spin multispin interacting model \cite{fendley2}.  The cases 
where $p=1$ and $N>2$ are the free-parafermionic Baxter models \cite{baxter1,baxter2, fendley1,baxter3,perk1,perk2,AB1,AB2}. The 
general cases where $p$ and $N$ are arbitrary was solved   in 
\cite{AP1,AP2}. This was done by extending the Fendley solution \cite{fendley2} for the fermionic case $N=2$ and $p=2$.  
The fermionic cases ($N=2$) with general values of $p$ are particular cases 
of free-fermion models defined in  frustration graphs 
\cite{network,circuits-pozsgay}. A more general related free-fermion model was also introduced recently 
\cite{fendley-pozsgay}.

Although the eigenenergies are 
exactly known for OBC the eigenfunctions are not known in a direct form. 
Interestingly the models show a phase diagram with a multicritical point with 
dynamical critical exponent $z= \frac{p+1}{N}$ \cite{AP1,AP2}, 
indicating that in general $z \neq 1$ and these models are not conformally invariant.
 Since most of the known critical quantum chains are conformally invariant, these 
exactly quantum chains  provide an interesting lab to explore  more general physical 
behaviors. 

 A natural question concerns the cases where $p+1=N$, and  $z=1$. Are the models 
conformally invariant in this case? Since conformal invariance imply, in the 
finite-size geometry, the existence of conformal towers in the eigenspectra, it is 
possible, from the exactly known eigenspectra, to verify this symmetry. It is 
interesting to mention that the ground-state energy of those non-Hermitian 
quantum chains ($N>2$) is real and the eigenenergies of the excited states appear 
in complex conjugated pairs.  Also these quantum chains have no chiral symmetry 
but in some cases they have a parity-time (PT) reversal symmetry. This PT symmetry however is 
broken since the eigenenergies appear in   complex conjugated pairs \cite{PT1}.

In Ref. \cite{AP3} it was shown that for $N>p$ there exist a set of $XX$ 
quantum chains with OBC that share the same quasi-energies 
that  give   the 
eigenenergies of the $Z(N)$ symmetric 
free quantum chains. The Hamiltonian besides two body interactions also contains 
$(p+1)$-multispin interactions. The equivalence happens up to overall degeneracies, mainly because the quasi-energies appear in  distinct combinations in both models. 
These $XX$ models have a $U(1)$ symmetry and are also non-Hermitian.  
Although sharing the  eigenspectrum with a parafermionic model they are described by fermionic 
operators through the Jordan-Wigner transformation.

The spectral equivalence among the $Z(N)$ and $XX$ Hamiltonians is valid for the OBC case. In the case of periodic boundary conditions (PBC) where the exact solution 
for the exact eigenspectrum of the $Z(N)$ model is unknown for $N>2$, the solution 
for the related $XX$ model is simple due to its free fermionic formulation. In this 
paper we explore the equivalence of the $Z(N)$ and $XX$ models to verify if indeed 
for the models where $N=p+1$ ($z=1$) the eigenspectrum is the one expected for 
quantum chains conformally invariant. 

The paper is organized as follows. In section II we present the free-particle models 
described in terms of fermionic and parafermionic operators. We present the models 
with $Z(N)$ symmetry and also the related $XX$ models with the larger $U(1)$ 
symmetry. In section III, we consider the solution of the models with OBC. The 
finite-size scaling of the eigenspectra of the models with OBC are given in section 
IV. In section V the eigenspectra of the periodic $XX$ model is studied. In section 
VI we present the entanglement properties of the $XX$ models with multispin 
interactions and PBC. Finally in section VII we draw our conclusion.

\section{Free-fermionic and free-parafermionic quantum chains}

Recently in \cite{AP1,AP2} it was shown that a large family of 
quantum chains have a free-particle spectra. These are Hamiltonians written as 
the sum of $M$ generators $\{h_i\}$,
\be \label{2.1}
H_M^{(N,p)} (\lambda_1,\ldots,\lambda_M) = -\sum_{i=1}^M \lambda_i h_i, 
\ee 
where 
$N=2,3,\ldots$,  and $p=1,2,\ldots$ are integers, and $\lambda_i$ are 
arbitrary coupling constants. The free-particle eigenspectra is a consequence 
of the $Z(N)$ exchange algebra, satisfied by the generators

\bea \label{2.2}
&&h_ih_{i+m} = \omega h_{i+m}h_i \quad\mbox{ for } \quad 1\leq	m \leq p;\quad  \omega = e^{i2\pi/N}, 
\nonumber \\
 && [h_i,h_j]=0 \mbox{ for } |j-i|>p,
\eea
with the closure relation 
\be \label{2.3} 
h_i^N = 1. 
\ee

Any representation of $\{h_i\}$ ($i=1,\ldots,M$) will have a free-particle 
eigenspectrum. The eigenenergies, apart from an overall representation dependent 
degeneracy (produced by zero modes), are given by 

\be \label{2.4}
E_{\scp{s_1,\ldots,s_{\mbar}}} = - \sum_{i=1}^{\mbar} \omega^{s_i} 
\varepsilon_i,
\ee
where 
\be\label{2.4p}
\overline{M} \equiv \mbox{int}\left(\frac{M+p}{p+1}\right)=\Big{\lfloor}{\frac{M+p}{p+1}}\Big{ \rfloor},
\ee
$s_i=0,1,\ldots,N-1$ and $\varepsilon_i$ ($i=1,\ldots,\mbar$) are the 
quasienergies of the pseudo-particles forming the eigenspectra. 

\begin{figure} [htb]
\centering
\includegraphics[width=0.45\textwidth]{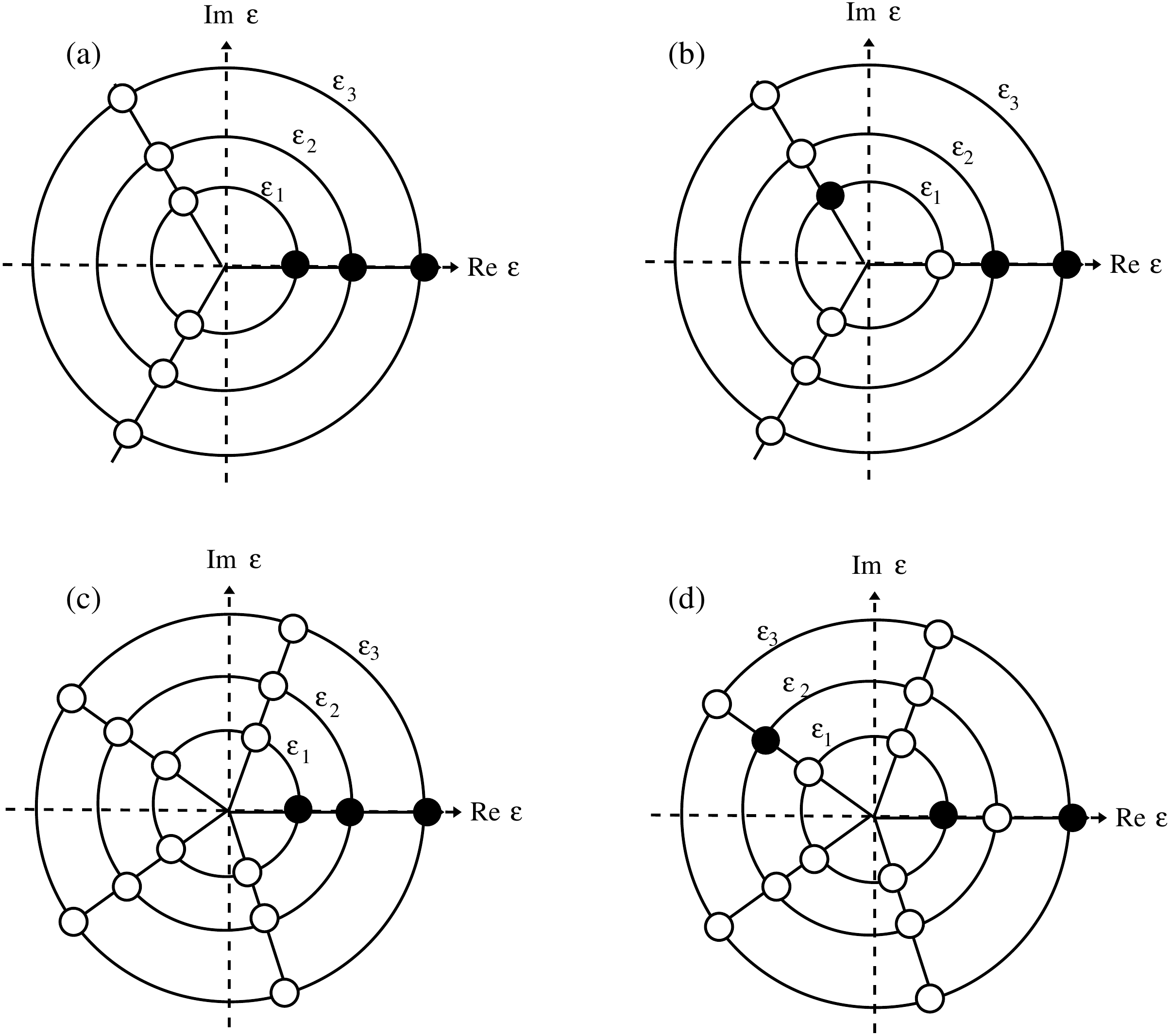}
\caption{  
Representation in the complex plane of the eigenenergies   \rf{2.4} with $\mbar=3$,  for 
the $Z(N)$ models with $N=3$ ((a) and (b)) and $N=5$ ((c) and (d)). The circles have the radius $\varepsilon_i$, and the possible values (open circles) are the intercepts with the $Z(N)$-circles. Each circle contributes with  {\it one and only one} of the possible $N$ 
intercepts (black circles).}\label{fig1}
\end{figure}

In Fig.\ref{fig1} we show schematically some eigenenergies for the $Z(3)$ model 
((a) and (b)) and for the $Z(5)$ model ((c) and (d)), in the case $\mbar=3$. Figs.\ref{fig1} (a) and (c) give a real eigenvalue and correspond in \rf{2.4} to the ground-state 
energy of the chains. The $3^3$ and $5^3$ energies for the $Z(3)$ and $Z(5)$ models are 
obtained by considering all the 3 or 5 allowed positions in the circles of 
radius $\epsilon_1$, $\epsilon_2$, and $\epsilon_3$, respecting a ``circle exclusion 
principle" that allows {\it{one and only one}} excitation in each circle. This is the $Z(N)$ 
parafermionic generalization of the standard $Z(2)$ Fermi-exclusion principle.

%%%%%%%%%%%%%%%%%
%%%%%%%%%%%%%%%%%
The pseudo 
energies $\varepsilon_i=1/z_i^{1/N}$ are obtained from the roots $z_i$ of a 
 polynomial $P_{\mbar}(z)$, generated  by the recurrence relation:

\be \label{2.5}
P_j(z) = P_{j-1}(z) -z\lambda_j^NP_{j-(p+1)}(z), j=1,2,\ldots ,
\ee
 with the initial condition

\be \label{2.14b} 
P_j(z) =1, \quad j\leq 0.
\ee
From \rf{2.4}, the representation of \rf{2.2} with $N=2$ will give us the 
Hamiltonian \rf{2.1} with a free-fermion eigenspectrum, while for $N>2$ we 
have  the $Z(N)$ free-parafermionic ones.

An interesting general representation of \rf{2.2} is given in terms of its 
independent words (word representation) (see Refs.\cite{AP1,AP2}). The Hamiltonian is given by 
\bea \label{2.7}
H_M^{(N,p)} &=& - \sum_{i=1}^p \lambda_i\left(\prod_{j=1}^{i-1} Z_j \right) X_i  \nonumber \\
&&- \sum_{i=p+1}^M \lambda_i\left( \prod_{j=i-p}^{i-1} Z_j\right) X_i,
\eea
where $Z$ and $X$ are the generalized $N\times N$ Pauli matrices satisfying 
\be \label{2.8}
XZ = \omega ZX, \quad X^N = Z^N=1, \quad Z^+ = Z^{N-1}.
\ee
The models contains $p+1$ multispin interactions and, except for $N=2$, are 
non-Hermitian.  We stress that, for $N>2$, the integrability is known only for OBC. For 
$N=2$ and $p=1$ we have the $M$-sites Ising-like model 
\be \label{2.9}
H_M^{(2,1)} =\lambda_1\sigma_1^x - \lambda_2\sigma_1^z\sigma_2^x + \cdots 
+ \lambda_M \sigma_{M-1}^z\sigma_M^x,
\ee
where $\sigma^x,\sigma^z$ are the standard spin-$\frac{1}{2}$ Pauli matrices.
Another representation for $N=2$, $p=1$ and $M$ odd is the standard quantum 
Ising chain with inhomogeneous couplings $\{\lambda_i\}$ and OBC
\be \label{2.10} 
H_{\mbox{\scp{Ising}}}= - \sum_{i=1}^L \lambda_{2i-1} \sigma_i^x 
-\sum_{i=1}^{L-1} \lambda_{2i}\sigma_i^z\sigma_{i+1}^z,
\ee
with $L=(M+1)/2$ sites. 
We can show that the Hamiltonians \rf{2.9} and \rf{2.10} share the same 
eigenenergies and degeneracies.

The case  where $N=2$ and $p=2$ in \rf{2.7} is the three-spin interacting 
Fendley model:
\bea  \label{2.11}
&&H_M^{(2,2)} = -\lambda_1 \sigma_1^x -\lambda_2\sigma_1^z\sigma_2^x 
-\lambda_3\sigma_1^z\sigma_2^z\sigma_3^x -  
\lambda_4 \sigma_2^z\sigma_3^z\sigma_4^x- \nonumber \\ 
&&\cdots  
-\lambda_{M-1}\sigma_{M-3}^z\sigma_{M-2}^z\sigma_{M-1}^x 
-\lambda_M \sigma_{M-2}^z\sigma_{M-1}^z\sigma_M^x.
\eea
The phase diagram of this model, in the homogeneous case 
$\lambda_1=\ldots,\lambda_M$  was studied in \cite{fendley2} and in 
\cite{ising-analogues}.

For $p=1$ and arbitrary $N$ the Hamiltonian \rf{2.7} is given by \rf{2.9} with 
the change $\sigma_i^x \to X_i$, $\sigma_i^z \to Z_i$ (satisfying \rf{2.8}).

Another representation in the case $p=1$, sharing the same eigenspectra is given by the known 
$Z(N)$ free-parafermionic Baxter model \cite{baxter1,fendley1,baxter2}
\be \label{2.12}
H_{\mbox{\scp{Baxter}}}^{(N)} = -\sum_{i=1}^L\lambda_{2i-1} X_i - \sum_{i=1}^{L-1} 
\lambda_{2i} Z_i Z_{i+1}^+, 
\ee
with $L= (M+1)/2$ sites.

Another special model  we are going to study in this paper is the $Z(3)$ 
version of the Fendley model, i.e., $p=2$ and $N=3$ in \rf{2.7}:
\bea \label{2.13}
&&H_M^{(2,3)} = -\lambda_1X_1 - \lambda_2Z_1X_2 -\lambda_3Z_1Z_2X_3 
-\lambda_4Z_2Z_3X_4 - \cdots \nonumber \\ 
&&-\lambda_{M-1}Z_{M-3}Z_{M-2}X_{M-1} 
-\lambda_MZ_{M-2}Z_{M-1}X_M.
\ee
It was shown for $N=2$, $p=1,2$ in \cite{fendley2} and for general $N,p$  in 
\cite{AP1,AP2} that the isotropic point $\lambda_i =1$ 
($i=1,\ldots,M$) is a multicritical point with the  energy gap vanishing as 
$\sim M^{-z}$, with the dynamical critical exponent value $z= (p+1)/N$. This 
means that in general $z\neq 1$ and the long-distance physics of the critical 
spin is not described by a conformal field theory (CFT). However for the 
special set of models where $p+1=N$, the dynamical critical exponent $z=1$ 
and the underlying field theory is relativistic and possibly conformal 
invariant.
This is the case for $p=1$ and $N=2$, that we recover the 
standard critical Ising quantum chains (\rf{2.10} with $\lambda_i=1$).
 However  for $N=p+1>2$ the time-evolution operator (the Hamiltonian) is non 
Hermitian. 

In the following sections we are going to compute the finite-size spectrum 
of these Hamiltonians with $N=p+1>2$ and verify the appearance of 
conformal towers as happens in the conformally invariant quantum chains.

\section{Generalized $XX$ quantum chains with multispin interactions}

In \cite{powerfull1,AP3} it was introduced a family of $XX$ quantum 
chains with two- and $N$-multispin interactions, with a free-fermion 
eigenspectrum whose quasienergies are the same as the $N$-multispin 
interacting $Z(N)$ models discussed in the previous section. The Hamiltonian 
is given by
\bea \label{3.1}
&&H_N^{XX} = -\sum_{i=1}^{L-1} \sigma_i^+\sigma_{i+1}^+  \nonumber 
 \\ 
&& - \sum_{i=1}^{L-N+1} \lambda_i^N \sigma_i^- \left(\prod_{j=i+1}^{i+N-2}
\sigma_j^z\right) \sigma_{i+N-1}^+,
\eea
where $\sigma^{\pm} = (\sigma^x \pm \sigma^y)/2$ are the standard 
raising/lowering spin-1/2 operators, $\{\lambda_i^N\}$ are the coupling 
constants and the lattice size is $L=M+N-1$.

It is interesting to observe that \rf{3.1} under the parity-symmetry (PT), where 
$i \to L-i+1$, the Hamiltonian transforms as $H_N^{XX} \to (H_N^{XX})^\dagger$, 
and from \cite{PT1}, the Hamiltonian although non-Hermitian can produce 
a unitary evolution. The Hamiltonian \rf{2.7} with $p=1$ has a PT symmetry \cite{bat-pt}, but not for general values of $p$. 
 
Differently from the $Z(N)$ models of last section the $XX$ Hamiltonians have a 
$U(1)$ invariance, since $\sum_{j=1}^{L} \sigma_j^z$ is a good quantum 
number. In the simplest case where $N=2$ the model recovers the dimerized 
version of the standard two-body $XX$ model:
\be \label{3.2}
H_2^{XX} (\{\lambda_i\}) = -\sum_{i=1}^{L-N+1} \sigma_i^+\sigma_{i+1}^- -
\sum_{i=1}^{L-N+1} \lambda_i^2 \sigma_i^-\sigma_{i+1}^+,
\ee
whose eigenspectrum is  well known to be related with the quantum Ising 
chain \cite{abb}.

The model \rf{3.1}, differently from \rf{2.1}-\rf{2.2}, is bilinear in therms of 
fermionic operators $\{c_i\}$ obtained from the Jordan-Wigner 
transformation \cite{lieb}
\be \label{3.3}
c_i = \sigma_i^-\prod_{j=1}^{i-1}(-\sigma_j^z),\quad
c_i^\dagger = \sigma_i^+\prod_{j=1}^{i-1}(-\sigma_j^z),
\ee
for $i=1,\ldots, L$,
that satisfy the anti-commutation relations 
\be \label{3.4}
 \{c_i,c_j^\dagger\} = \delta_{i,j},\quad \{c_i,c_j\}=\{c_i^{\dagger},c_j^{\dagger}\}=0.
\ee
Since $\sigma_j^z= 2c_j^{\dagger}c_j -1$, the $U(1)$ symmetry is translated 
into the conservation of the total number of fermions 
$N_F = \sum_{i=1}^{L} c_i^{\dagger}c_i$, and the $z$-magnetization of the XX 
multispin model is given by 
\be \label{3.4a}
m_z= \sum_{j=1}^{L} \sigma_j^z = 2N_F - L.
\ee
 In terms of $\{c_i\}$ the Hamiltonian \rf{3.1} has the bilinear form 
\be \label{3.5}
H = -\sum_{i,j=1}^{L} c_i^\dagger \mathbb{A}_{i,j} c_j, 
\ee
where 
\be \label{3.6}
\mathbb{A}_{i,j} = \delta_{j,i+1} + \lambda_j^N\delta_{j,i+1-N},
\ee
are the elements of the matrix formed by the hopping coupling constants. 

The matrix $\mathbb{A}$ for $N>2$ is non-symmetric, nevertheless can be 
diagonalized:
\be \label{3.7}
H = -\sum_{k=1}^{L} \Lambda_k \eta_k^{\dagger} \eta_k,
\ee
through the canonical transformation $\{c_i,c_i^{\dagger}\} \to 
\{\eta_k,\eta_k^{\dagger}\}$
\be  \label{3.8}
\eta_k = \sum_i^{L}\mathbb{L}_{i,k} c_i, \quad 
\eta_k^\dagger = \sum_i^{L}\mathbb{R}_{i,k} c_i^\dagger,
\ee
where in \rf{3.7} $\Lambda_k$ are the eigenvalues of $\mathbb{A}$ and 
$\mathbb{L}_{i,k}$,$\mathbb{R}_{i,k}$ are the components  of the left and 
right eigenvectors with the normalization  $\mathbb{R}\mathbb{L}^T=\id$,
 respectively. 

From \rf{3.7} the eigenenergies of $H_{N}^{XX}$ have the free-fermion 
structure
\be \label{3.9}
H = -\sum_{k=1}^{M+N-1} s_k \Lambda_k; \quad s_k=0,1.
\ee
The quasienergies $\{\Lambda_k\}$ are obtained from the roots of 
$\mbox{det}(\mathbb{A}- 
\Lambda_k \id)=0$. Apart from the zero modes they are given by 
$\Lambda_k = 1/z_k^{1/N}$, where $z_k$ are the roots ($P_M^{(N)}(z_k)=0$) of 
the characteristic polynomial
\be \label{3.10}
P_M^{(N)}(z) \equiv \mbox{det}(\id -z\mathbb{A}). 
\eea
From the Laplace cofactor's rule for determinants, these polynomials 
satisfy the recurrence relations:
\be \label{3.11}
P_M^{(N)}(z) = P_{M-1}^{(N)}(z) -z\lambda_M^N P_{M-N}^{(N)}(z),
\ee
with the initial condition 
\be  \label{3.12}
P_M^{(N)}(z) = 1, \mbox{for } M\leq 0,
\ee
and $z_k=(1/\Lambda_k)^N$.
Comparing \rf{2.5}-\rf{2.14b} and \rf{3.11}-\rf{3.12} we see that the 
polynomials  $P_M^{(N)}(z)$ are the same as those fixing the eigenspectra 
of the $Z(N)$ multispin chains with $N=p+1$. Namely, the same 
roots $\{z_k\}$ that 
give the quasienergies $\epsilon_k= 1/z_k^{1/N}$ of the $Z(N)$ 
free-parafermionic multispin models also give the ones of the 
$XX$ chains with $N$-multispin interactions i. e.,
\be \label{3.13}
\Lambda_{j,i} = e^{\frac{2\pi}{N}j} \epsilon_i,
\ee
where $i=1,\ldots,
\lfloor\frac{L}{N}\rfloor$ and  $j=0,1,\ldots,N-1$.
Since the dimension of $H_N^{XX}$ is $2^{L}$ and the total number of 
nonzero quasienergies $\{\epsilon_i\}$ is $L$, we should have 
$N_z=L -N\lfloor\frac{L}{N}\rfloor$ zero modes, producing a 
$2^{N_z}$-global degeneracy of the whole eigenspectra of the Hamiltonian
\be \label{3.14}
E_{\{s_{i,j},r_{i,j}\}} =
-\sum_{i=1}^{\inpar{\frac{L}{N}}} 
\left( \sum_{j=0}^{N-1} r_{i,j}
\omega^{s_{i,j}}\right) \varepsilon_i, 
\ee
where for each $i=1,\ldots,\inpar{\frac{L}{N}}$,   
we have a possible  $s_{i,j}=0,1,\ldots,N-1$ and 
$r_{i,j}=0,1$.

The schematic representations, similarly as shown in Fig.~1 for the $Z(3)$ 
model, are in circles of radius $\epsilon_i$, but now in a given circle 
we have $2^N$ possible occupations of pseudo-particles. All the eigenenergies 
represented in Fig.~1 for the $Z(3)$ model are also presented in the 
3-multispin $XX$ model, including the ground-state energy. The eigenlevels 
shown in Fig.~2 are present in the $XX$ multispin model, but not in the 
corresponding $Z(3)$ $p=2$ model with $\mbar =2$, since they do not obey the circle 
exclusion  constraint. The circle exclusion  in the $Z(N)$ models 
is not a constraint for these $U(1)$ $XX$ models.  
%%%%%%%%%%%%%%%%%
\begin{figure} [htb]
\centering
\includegraphics[width=0.45\textwidth]{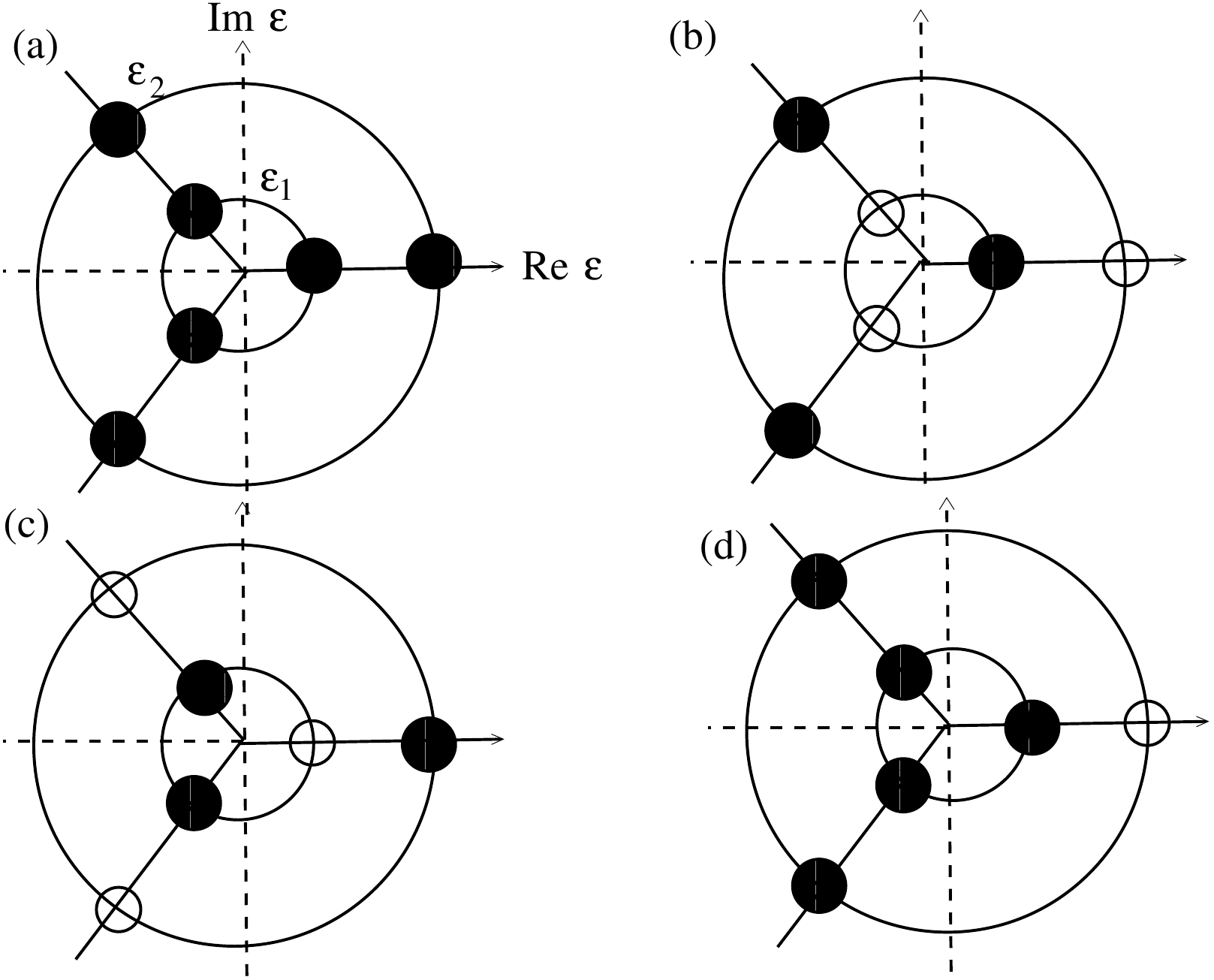}
\caption{  
Representation of configurations that are present in the $N=3$-multispin $XX$ quantum chains \rf{3.14} but not 
present in the multispin 
$Z(3)$ quantum chain \rf{2.4}, in the case $\mbar=2$. 
The circles have the radius $\varepsilon_i$, and the possible values (open circles) are the intercepts with the $Z(N)$-circles. 
Distinct from the configurations in Fig.~1, each circle may have multiple contributions (black circles), and do not satisfy the 
$Z(N)$-circle exclusion constraint.}\label{fig2}
\end{figure}
%%%%%%%%%%%%%%%%%

\section{The eigenspectra of the $N$-multispin interaction models with open boundary conditions (OBC)}

In this section we calculate the eigenenergies of the model with $N$-multispin interactions 
 with $Z(N)$ and $U(1)$ symmetries presented in section II and III, with OBC. 
We restrict ourselves to the isotropic couplings case where 
$\lambda_i = \lambda$ ($i=1,\ldots,M$). In this case the roots 
$z_i(\lambda)$ of the polynomial \rf{2.5} that fix the quasienergies obey 
$z_i(\lambda) = z_i(1)/\lambda^N$ and the quasienergies 
$\epsilon_i(\lambda) = \lambda\epsilon_i(1)$. This imply that the 
eigenspectra of the Hamiltonian satisfy:
\be \label{4.0}
H_{\mbox{\scp{OBC}}} (\lambda) = \lambda H_{\mbox{\scp{OBC}}}(1).
\ee
This also follows directly from \rf{2.1} in the case of the $Z(N)$ 
multispin models. In the case of the  $N$-multispin $XX$ model it is not direct 
but in section 5 we give a canonical transformation of the spin variables 
that also show \rf{4.0} directly. 

At their isotropic point the models are critical with a dynamical critical 
exponent $z=1$. They are given by \rf{2.7} with $p=N+1$ 
in the $Z(N)$ case and by \rf{3.1} in the $XX$ case.

The multispin $Z(N)$ Hamiltonian  with $M$ sites have the energies
\be \label{4.1}
E_{\scp{\{s_1,\ldots,s_{\mbar}\}}} = -\sum_{i=1}^{\mbar} \omega^{s_i} 
\epsilon_i, \quad s_i=0,1,\ldots,N-1,
\ee
while the $XX$ model with $L=M+N-1$ sites have the energies
\be \label{4.2}
E_{\scp{\{t_{i,j},r_{i,j}}\}} = -\sum_{i=1}^{\mbar} \left(
\sum_{j=0}^{N-1}r_{i,j}\omega^{t_{i,j}}\right) \epsilon_i,
\ee
with $r_{i,j}=0,1$, $t_{i,j}=0,1,\ldots,N-1$, and from \rf{2.4p} 
$\mbar = \left\lfloor \frac{L}{N} \right\rfloor$ .

The pseudo-energies $\epsilon_i$ ($i=1,\ldots,\mbar$) are the same for both 
models and can be evaluated from the roots $z_i$ of the polynomial \rf{3.11},
since $\epsilon_i = 1/z_i^{1/N}$, or directly from the diagonalization of 
the hopping matrix $\mathbb{A}$ given by \rf{3.6}. In the critical region the 
low-lying quasiparticles $\{\epsilon_i\}$ give us the relevant excitations 
and we should expect that as $M\to \infty$ they should vanish. This means  
that in this limit, the large roots of the polynomial \rf{3.11} should diverge. 

The roots $\{z_i\}$ are all real and the eigenenergies of \rf{4.1} and \rf{4.2} are in general complex, thanks to the non-hermicity of the Hamiltonian 
\rf{2.7}  and \rf{3.1}. 

For convenience we order the quasienergies 
$\epsilon_1 <\epsilon_2 < \cdots < \epsilon_{\mbar}$, while the eigenenergies 
of the Hamiltonian we order in  increasing order of their real part:
\be \label{4.3}
\Re ({E_0}) \leq \Re({E_1}) \leq \Re(E_2) \leq \cdots .
\ee
The lowest eigenenergy, that give us the ground-state energy is real and is given 
by
\be \label{4.4}
E_0(M) = - \sum_{i=1}^{\mbar} \epsilon_i.
\ee

The ground-state energy can be derived using the results of \cite{AP2}, and is 
given by
\bea \label{4.4a}
&&e_{\infty} = \lim_{M\to \infty} \frac{E_0}{M} = 
-\frac{1}{N\pi} \int_{0}^{\pi}
\frac{\sin x}{\sin^{\frac{1}{N}} (\frac{x}{N}) 
\sin^{\frac{N-1}{N}}
 (\frac{(N-1)x}{N})} \nonumber \\= 
&&-\frac{N\sin(\frac{\pi}{N})}{(N-1)\pi},
\eea
with the values
\beq  \label{4.44a}
-\frac{2}{\pi} , -\frac{3\sqrt{3}}{4\pi}, -\frac{2\sqrt{2}}{3\pi}, -\sqrt{\frac{5-\sqrt{5}}{8\pi}},
\eeq
for $N=2-5$, respectively.
 
In order to verify the conformal invariance of the models at their 
critical points, we are going to explore the consequences of the underlying 
conformal symmetry in the finite-size eigenspectrum of the quantum chains 
with OBC.

The finite-size amplitudes of the excited states 
will give us the surface exponents of the model. To each surface exponent 
$x_s$ of the infinite system \cite{cardy1986b} we should expect,
at the critical point  a tower of eigenenergies 
\be \label{4.6}
\Re (E_s(M,r)) = E_0(M) + \frac{\pi v_s (x_s +r)}{M} + o(M^{-1}), 
\ee
where $r=0,1,2,\ldots$.

Since the pseudo-energies $\{\epsilon_i= 1/z_i^{1/N}\}$ in \rf{4.1} and \rf{4.2}
 depend on the roots $\{z_i\}$ of the polynomials \rf{2.5} or \rf{3.11} it is 
convenient to observe their asymptotic finite-size dependences. In the cases 
where the model is critical we should expect that the large roots diverge
with the order of the polynomial. In \cite{powerfull1} it was introduced  a method 
that allow us to evaluate these large roots for huge lattice sizes 
($\sim 10^9$) by using standard quadruple-precision numerical calculations. The 
coefficients of the polynomial have quite small and large numbers. The method works if we have a good initial guess for the roots. For the largest root, the 
Laguerre bound (see corollary 6.2.4 on \cite{laguerre}) is the initial value. After the 
evaluation of the largest root we produce good initial guesses for the other 
roots by exploring the size dependence of the largest root. With this 
procedure we evaluate the largest $10-12$ roots for polynomials up to the 
order $M=10^9$. 
The method was also tested in the random formulation of free-fermion models ($N=2$), with $p=1$ \cite{powerfull1} and more recently for $p=2$ models \cite{randonp2}.

%parei
The prediction \rf{4.6} indicates  the leading behavior for 
the roots of the polynomial $P_M^{(N)}(z)$ given in \rf{3.11}-\rf{3.12}:
\be \label{4.7}
\frac{1}{z_i^{1/N}} 
= \epsilon_i = \pi \frac{A_i^{(N)}}{M} = 
\pi\frac{A^{(N)}} {M} (x_s^{(N)} + i-1), 
\ee
for $i=1,2,\ldots$. 
 The amplitude $A^{(N)}$ is proportional to the sound velocity
and 
$x_s^{(N)}$ is a 
surface exponent.

Our numerical solutions for polynomial roots with $M$ up to $10^9$ 
corroborate the conformal invariance prediction \rf{4.7} for the roots. We 
conjecture the following exact values for $N=2, 3,4$ and $6$ 
\be \label{4.8}
A^{(2)} = 2, \quad A^{(3)}= 2\sqrt{3}, \quad A^{(4)} = 4\sqrt{2}, A^{(6)}=12,
\ee
and the numerical value
\be \label{4.9}
A^{(5)} = 8.506283,
\ee
for $N=5$. The surface exponent values depend on the particular 
sequence of lattice sizes ($\mod(M,N)$  fixed), used to obtain the bulk limit $M\to \infty$. The results we obtain for $N=2$ and $N=3$ are
\bea \label{4.10}
&&x_s^{(2)}= 1, \mbox{ (mod($M,2$)=0 or 1)},  \nonumber \\
&&x_s^{(3)} = 7/6 \mbox{ (mod($M,3$)=0)}, x_s^{(3)} = 1/6  \mbox{ (mod($M,3$)=1)}, \nonumber \\
&&x_s^{(3)} = 5/6 \mbox{ (mod($M,3$)=2)},
\eea
while for $N=4$, 
\bea \label{4.12}
&&x_s^{(4)} = 5/4 \mbox{ (mod($M,4$)=0)}, x_s^{(4)} =1/2 \mbox{ (mod($M,4$)=1)}; \nonumber \\
&& x_s^{(4)} = 3/4 \mbox{ (mod($M,4$)=2)}, x_s^{(4)}=1 \mbox{ (mod($M,4$)=3)}.
\eea
In order to illustrate the numerical results we show in Table~I the ratios
 $A_i^{(N)}/A^{(N)}$ a  conformal tower of the model with $N=2-5$ , for $M=10^9-N-1$ (mod$(M,N)=0$).
\begin{table*}[t]
\centering
\begin{tabular}{|c|c|c|c|c|}
\hline 
$i$ & $A^{(2)}(i)/A^{(2)}$ & $A^{(3)}(i)/A^{(3)}$ &  $A^{(4)}(i)/A^{(4)}$ & $A^{(5)}(i)/A^{(5)}$\\ \hline  \hline
1 & 1 & 1.1669439 & 1.24987070 & 1.333553  \\ \hline 
2 & 2 & 2.1666355 & 2.24998504 & 2.333366 \\ \hline 
3 & 3 & 3.1666632 & 3.24997209 & 3.333367 \\ \hline 
4 & 4 & 4.1666613 & 4.24995483 & 4.333357 \\ \hline 
5 & 5 & 5.1666589 & 5.24993348 & 5.333337\\ \hline 
6 & 6 & 6.1666562 & 6.24990804 & 6.333328 \\ \hline 
12 & 12 & 12.1666314 & 12.24966964 & 12.332937 \\ \hline 
\end{tabular}
\caption{Examples of estimates for the ratios $ A_i^{(N)}/A^{(N)}$ for some 
conformal towers of 
the models with $N=2-5$. 
The ratios are the ones of the lattice sizes $M=10^9-N-1$, where mod$(M,N)=0$.}
\label{table1}
\end{table*}

The finite-size dependence of the mas gaps \rf{4.6} of the $Z(N)$ and $XX$ 
models with $N$-multispin interactions can be obtained directly from the 
relations \rf{4.7}-\rf{4.9}.

Let us consider initially the $Z(N)$ free-parafermionic models.
The ground-state energy (real) is obtained (see \rf{4.4}) by considering all
 the roots in the branch $\omega^0=1$ ( as in Fig.~3a). A sequence of mass gaps with lower 
real part is obtained by changing a root $\epsilon_i$ ($i=1,2,\ldots$) 
in the ground state (see Fig.~3(a))  to the $\omega^1 = e^{i2\pi/N}$ or 
$\omega^{N-1} = e^{-i2\pi/N}$ branches (see  Figs.~3b,c)
\be \label{4.14}
\Re(G_i) = \Re (\epsilon_i - \omega \epsilon_i) = 
(1-\cos(\frac{2\pi}{N}))\epsilon_i, 
\ee
($i=1,2,\ldots$).
The relations \rf{4.8} and \rf{4.9} give us the sound velocity, in \rf{4.6}, 
for the models
\bea \label{4.15}
&& v_s^{(2)}=2A^{(2)}=4, \nonumber \\
&&v_s^{(3)} = 3A^{(3)}/2 = 3\sqrt{3}, \quad v_s^{(4)}= 4\sqrt{2}, \nonumber \\
&& v_s^{(5)} = A^{(5)} (1- \cos(2\pi/5)) \approx 0.69098300, \nonumber \\
&& v_s^{(6)} = A^{(6)} (1-\cos(2\pi/6)=1/2,
\eea
and the conformal towers 
\be \label{4.17}
x_s^{(N)} + i-1; \quad i=1,2,\ldots,
\ee
given in \rf{4.10} and \rf{4.12}.

The conformal anomaly $c$ is also predicted from the leading finite-size 
behavior of the ground-state energy $E_0(L)$. At a critical point, should 
behave asymptotically as \cite{blote1}
\be \label{4.5}
\frac{E_0}{L} = e_{\infty} + f_s L - \frac{\pi c v_s }{24L}  + o(L^{-1}),
\ee 
where $e_{\infty}$ and $f_s$ are, respectively, the bulk limits of the 
ground-state and surface energy per site, $v_s$ and $c$ are the sound velocity 
and the conformal anomaly. The use of the above prediction is not simple 
because $L$ is the effective number of sites of the space discretization 
of the underlying conformal field theory, and the relation with the number $M$ 
in our models, except for the case $N=2$, are not direct. To better explain 
this point we show below the expansions \rf{4.5} up to order $o(M^{-2})$ for the 
cases $N=2,3$. For the case 
$N=2$:
\bea	 \label{4.5b}
&&E_0 = -\frac{2}{\pi}M + (1-\frac{4}{\pi}) + \frac{\pi}{6M}, 
 \mbox{Mod($M,2$)=0}, \nonumber \\
&&E_0 = -\frac{2}{\pi}M + (1-\frac{4}{\pi}) - \frac{\pi}{12M}, 
 \mbox{Mod($M,2$)=1},
\eea
while for $N=3$:
\bea	 \label{4.5c}
&&E_0 = -\frac{3\sqrt{3}}{4\pi}M -0.46909 +\frac{1.9615}{M}, 
 \mbox{Mod($M,3$)=0}, \nonumber \\
&&E_0 = -\frac{3\sqrt{3}}{4\pi}M -0.46909 -\frac{0.5610}{M}, 
 \mbox{Mod($M,3$)=1}, \nonumber \\
&&E_0 = -\frac{3\sqrt{3}}{4\pi}M -0.46909 +\frac{1.6985}{M}, 
 \mbox{Mod($M,3$)=2}. \nonumber \\
\eea
The expansion for $N=2$ was calculated analytically and the ones for $N=3$ was obtained by a cubic fitting, 
considering $60<M<600$. The expansion in \rf{4.5b} with $M$ odd recovers \rf{4.5} if we identify 
the Ising quantum chain representation \rf{2.10} with $M=2L-1$, $v_s=2$ and $c=1/2$. 
This is not the case for the expansion fort $N=2$ and $M$ even, where the $O(1/M)$ term 
is positive instead of negative as in \rf{4.5}. The expansions for the $N=3$ cases also 
give us terms that we cannot compare directly with \rf{4.5}. We leave the conformal anomaly 
calculations for the next section where  we consider the periodic lattices. 

We have also, in the $Z(N)$ models, the excited states formed by replacing
$\ell$ ($\ell=2,3,\ldots$) quasienergies in the branch $\omega^0$ by 
quasienergies in the branches $\omega^1$ or $\omega^{N-1}$ (see Fig.~4). 
%%%%%%%%%%%%%%%%%
\begin{figure} [htb]
\centering
\includegraphics[width=0.35\textwidth]{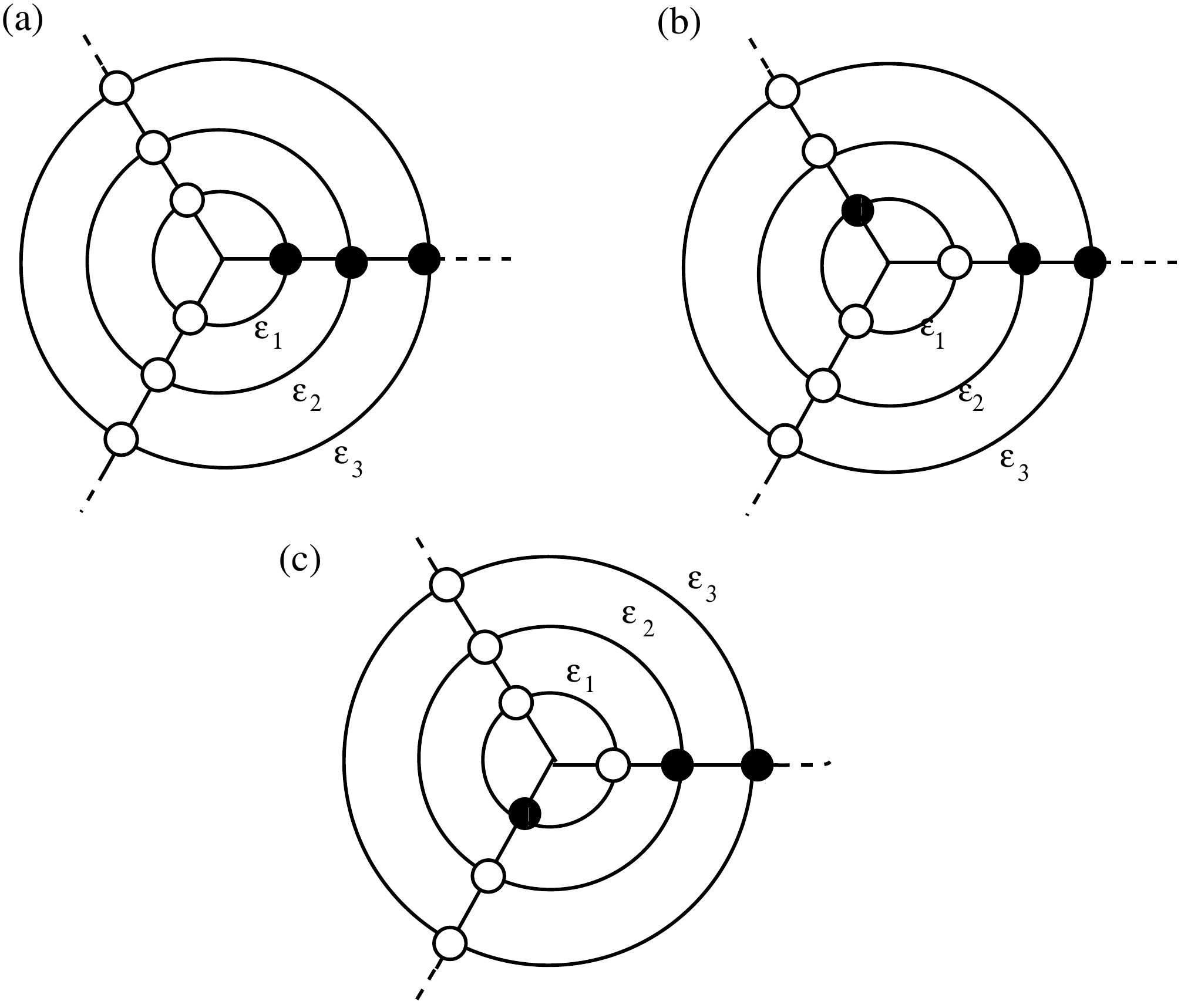}
\caption{  
Representation in the complex plane of the eigenenergies   for 
the $Z(N)$ multispin model with $N=3$. 
The configuration of the ground state energy is (a), and (b),(c) are the configurations 
that produce the lowest gaps \rf{4.14}.}
\end{figure}
%%%%%%%%%%%%%%%%%
From \rf{4.7} and \rf{4.17} the mas gaps associated to these states give us 
the conformal dimensions 
\be \label{4.18}
\ell x_s^{(N)}  + j,
\ee
with $\ell,j \in \mathbb{Z}$ and $\ell \geq 1$ and $j\geq \ell(\ell+1)/2$.

For $N>3$ we can produce several other conformal dimensions since we may 
consider the eigenstates where the particles are in larger number of 
branches $\omega^n$ ($n=0,\ldots,N-1$).

For the case of the $XX$ models with $N$-multispin interactions, we have the 
same quasienergies $\epsilon_i$, considered in the $Z(N)$-parafermionic models, 
but their possible  combinations  are not restricted to the $Z(N)$ circle exclusion.
This imply that a given quasienergy $\epsilon_i$ can appear up to $N$ times 
($\omega^n \epsilon_i, n=0,\ldots,N$) in a given eigenenergy of the 
Hamiltonian. The model has now a $U(1)$ symmetry, and we can separate the 
associated eigenvector space according to its magnetization, or 
equivalently, to the number $N_F$ of fermionic quasienergies. The 
magnetization is given by \rf{3.4a}.
%\be \label{4.19}
%m_z = \sum_{j=1}^{L} \sigma_j^z = 2N_F - L.
%\ee
For $N\leq4$ the ground-state is formed by taking all the 
$\mbar =\lfloor \frac{L}{N} \rfloor$ roots a single time in the 
branch $\omega^0=1$. It belongs to the sector where $m_z^{(0)} = 
2\mbar - L$. This energy coincides with the ground-state energy of the 
corresponding $Z(N)$ parafermionic quantum chain. Actually all the 
eigenenergies we consider previously in the $Z(N)$ model are also present in 
the sector with magnetization $m_z^{(0)}$, giving us the same sound velocity 
and conformal dimensions given in 
\rf{4.10}-\rf{4.17}. 
The absence of the $Z(N)$-circle exclusion gives additional conformal 
dimensions, inside the $m_z^{(0)}$ sector. They are formed by neglecting an 
arbitrary number of roots forming the ground state (branch $\omega^0$) and 
inserting them in the other branches, keeping the number of fermions $N_F$ fixed
 (some examples of excitations are shown in  Fig.~4). 

The conformal dimensions coming from the eigensectors with other 
magnetizations are obtained by neglecting  and inserting distinct number of particles in the ground-state pseudo-particles configuration. It is simple to 
verify that some of the produced gap will give the same conformal dimensions 
\rf{4.15}-\rf{4.18}, but with a distinct sound velocity that depends on the 
particular magnetization sector. This means that distinct from the 
$Z(N)$ parafermionic model, the $XX$ multispin model in the bulk limit, is a 
combination of distinct theories with unequal sound velocities.

For $N>4$ the ground-state energy of the related $Z(N)$ parafermionic quantum 
chain is in the sector $m_z^{(0)}$, but in the $XX$ model it is an excited 
state. The energy with lowest real part in the $XX$ model is obtained by 
adding all the roots in the branches $\omega^{\pm\ell}$ ($\ell=0,1,\ldots,\lfloor \frac{N-1}{2}\rfloor$) (se Fig.~5), and gives 
\be \label{4.20} 
E_0 = -(1 + 2\sum_{\ell=1}^{\lfloor \frac{N-1}{4}\rfloor} \cos(\frac{2\pi}{N}\ell))
\sum_{i=1}^{\mbar} \epsilon_i.
\ee 

%{\bf Falar sobre einfty e anomalia conforme para o caso open - veja file estrela}

%%%%%%%%%%%%%%%%%
\begin{figure} [htb]
\centering
\includegraphics[width=0.45\textwidth]{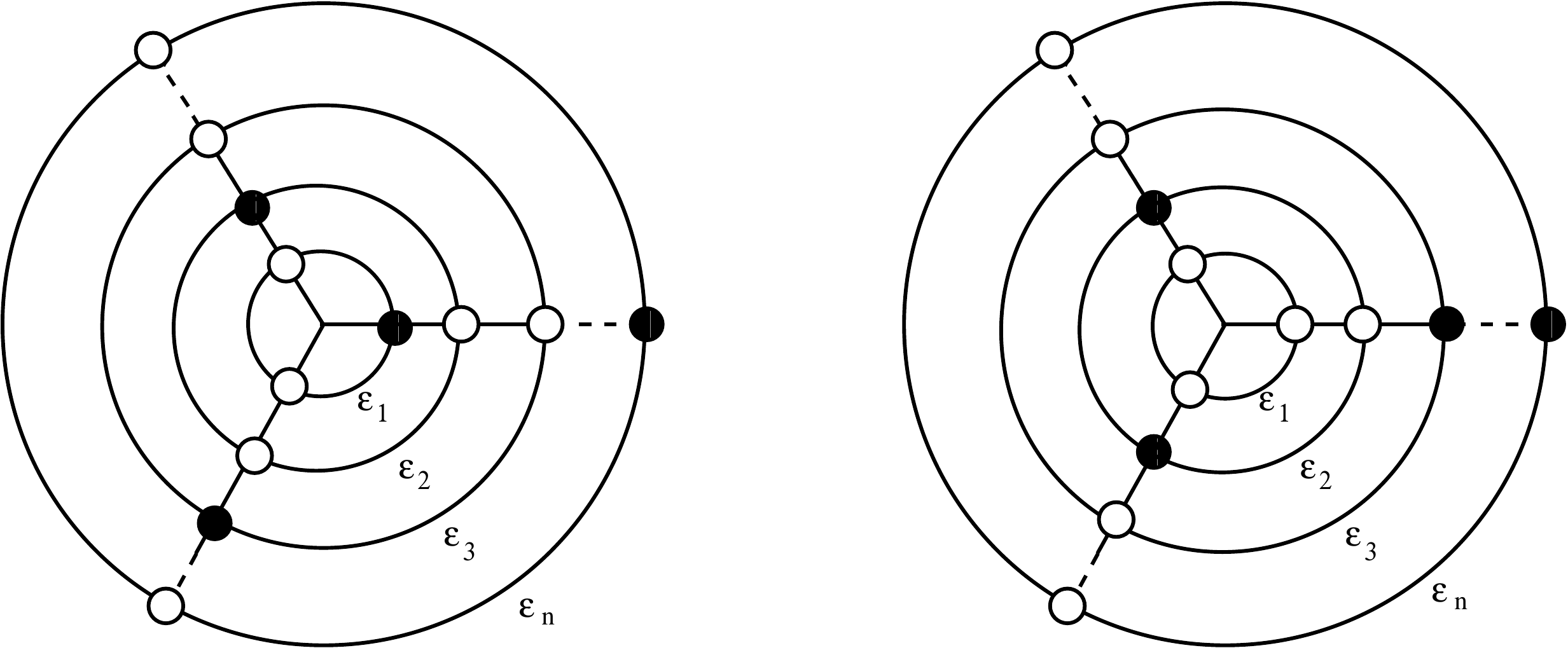}
\caption{  
Representation in the complex plane of the energies of some excited states for 
the $N=3$ $XX$ multispin model.} \label{fig4}
\end{figure}
%%%%%%%%%%%%%%%%%
%%%%%%%%%%%%%%%%%
\begin{figure} [htb]
\centering
\includegraphics[width=0.2245\textwidth]{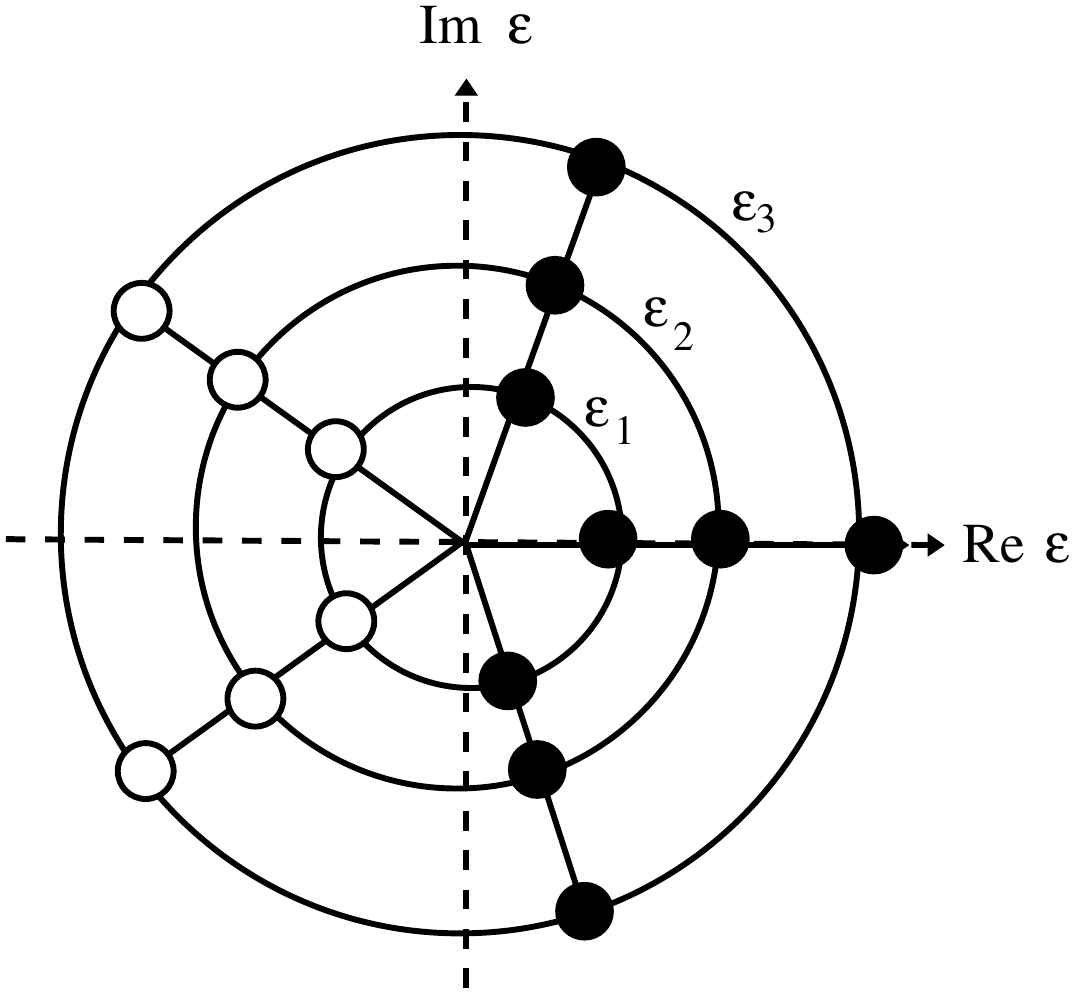}
\caption{  
Representation in the complex plane of the ground-state eigenenergy of the $N=5$ 
multispin $XX$ model.} \label{fig5}
\end{figure}
%%%%%%%%%%%%%%%%%

\section{The $N$-multispin $XX$ models with periodic boundary conditions (PBC)}

The eigenspectral equivalence among the $Z(N)$ multispin models and the $XX$ multispin quantum chains only holds in the case of OBC.
 Previous numerical studies \cite{AB2} of the $Z(3)$ parafermionic Baxter model 
($p=1,N=3$ in \rf{2.7}) show us that the quantum chain with PBC have quite 
distinct properties from the chain with OBC. The energy per site of the ground-state energy of the different boundary conditions are distinct. We should 
expect a similar effect for the more general $Z(N)$ parafermionic quantum 
chains with multispin interactions \rf{2.7}. 

In this section we are going to study the $N$-multispin $XX$ models with 
isotropic couplings and PBC. The Hamiltonian has $L=M+N-1$ sites and is given by
\be \label{5.1}
H_{\scp{PBC}} (\lambda) = H_{\scp{OBC}}(\lambda) + H_s^{(L)} + H_s^{(R)}, 
\ee
where 
\bea \label{5.2} 
&&H_{\scp{OBC}} (\lambda) = \sum_{i=1}^{L-1} \sigma_i^+\sigma_{i+1}^- \nonumber \\ 
&+& \lambda^N \sum_{i=1}^{L-N+1} \sigma_i^- \left(\prod_{j=i+1}^{i+N-2} 
\sigma_j^z \right) \sigma_{i+N-1}^+,
\eea
is the $XX$ Hamiltonian with OBC and 
\be \label{5.3}
H_s^{(L)} = \sigma_{L}^+\sigma_1^-,
\ee
\bea \label{5.4}
H_s^{(R)} &=& \lambda^N \sum_{\ell=1}^{N-1} \sigma_{L+\ell-N +1}^- 
\nonumber \\
&& \times \left( 
\prod_{k=L+2+\ell-N}^L \sigma_k^z\right) \left(\prod_{t=1}^{\ell -1} 
\sigma_t^z\right) \sigma_{\ell}^+ 
\eea
 are the left and right surface terms.

It is interesting to consider the site-dependent canonical transformation:
\be \label{5.5}
\sigma_i^{\pm} \rightarrow (\lambda)^{\mp(i-1)} \sigma_i^{\pm};\quad \sigma_i^z 
\rightarrow \sigma_i^z \quad (i=1,\ldots,L),
\ee
that transform the Hamiltonian \rf{5.1} into:
\bea \label{5.6}
H_{\scp{PBC}}(\lambda) &=& \lambda H_{\scp{OBC}}(1) \nonumber \\
&+& \frac{1}{\lambda^{L-1} }
H_s^{(L)} + \lambda^{L+1} H_s^{(R)} (1).  
\eea
This result tell us that the spectral symmetry \rf{4.0} only holds in the 
OBC case, and for $\lambda \ne 1$ we may expect distinct behavior for the PBC. 
This happens 
even in the bulk limit since the surface terms gave contributions that are 
exponentially large with the system's size.

The $U(1)$ symmetry of the model allow us to split the associated 
vector space of the Hamiltonian \rf{5.1} into  sectors labelled by the 
$z$-magnetization $m_z$.
On each sector, where the number of fermion is $N_F=(m_z+L)/2$,  
we can perform the Jordan-Wigner transformation given in 
section III, and the Hamiltonian \rf{5.1} takes the form:
\be \label{5.8a}
H_{\scp{OBC}} (\lambda) = -\left( \sum_{i=1}^{L-1} c_i^{\dagger}c_{i+1} + 
\lambda^N \sum_{i=1}^N c_{i+N-1}^{\dagger} c_i \right),
\ee
\bea \label{5.8b}
&&H_s^{(L)} = - (-)^{L+N_F+1}c_L^{\dagger}c_1, \nonumber \\ 
&&H_s^{(r)}= - (-)^{L+N_F +1} \sum_{\ell=1}^{N-1} c_{\ell}^{\dagger} c_{M+\ell},
\ee
so that 
\be \label{5.9a}
H_{\scp{PBC}} (\lambda) = -\left(\sum_{i=1}^L c_i^{\dagger}c_{i+1}, 
 + \lambda^N\sum_{i=1}^L c_{i+N-i}^{\dagger} c_i \right),
\ee
with the boundary condition
\be \label{5.9b}
c_{L+\ell} = (-)^{L+N_F+1} c_{\ell} \quad  (\ell=1,2,\ldots).
\ee
The fermionic model is periodic or antiperiodic depending if $L+N_F+1$ is even or odd, respectively. 

In order to diagonalize \rf{5.9a} we perform the Fourier transformation 
$\{c_i\} \to\{\eta_k\}$, where 
\be \label{5.10a}
\eta_k=\frac{1}{\sqrt{L}} \sum_{j=1}^L e^{ikj}c_j, \quad c_j = 
\frac{1}{\sqrt{L}}\sum_{\{k\}} e^{-ikj} \eta_k. 
\ee
It follows from the algebraic relations of $\{c_j\}$ \rf{3.4} that $\{\eta_j\}$ 
are also fermionic operators:
\be \label{5.10b}
\{\eta_k,\eta^{\dagger}_{k'}\} = \delta_{k,k'}, \quad
\{\eta_k,\eta_{k'}\}=0.
\ee
Inserting \rf{5.10a} in \rf{5.9b} we obtain the sets
\be \label{5.10c}
k_j= 
\begin{cases}
\frac{2\pi j}{L}, & \text{if $L+N_F+1$ even} \\
\frac{2\pi (j+\frac{1}{2})}{L}, & \text{if $L+N_F+1$ odd}
\end{cases},
\ee
and $\{k_j\}$ are chosen inside  one of the Brillouin zones, e.g., 
$-\pi < k_j \leq \pi$.

The Hamiltonian \rf{5.6} in terms of $\{\eta_k\}$ is diagonal
\be \label{5.11}
H=\sum_{\{k_j\}} \epsilon(k_j) \eta_{k_j}^{\dagger}\eta_{k_j}
\ee
where
\be \label{5.12}
\epsilon(k_j) = - (e^{-ik_j} + \lambda^N e^{i(N-1)k_j}),
\ee
and the momentum of a given state is 
$P=\sum_{\{k_j\}} \eta_{k_j}^{\dagger} \eta_{k_j}$.

As in the OBC we order the eigenvalues in increasing order of their real part. 
The effective dispersion relation is:
\be \label{5.13} 
\Lambda(k) = \Re (\epsilon(k)) = -(\cos{k} +\lambda^N\cos [(N-1)k]).
\ee
This is also the dispersion relation of an extended Hermitian $XX$ model, considered in
\cite{tit}.
The momentum and the real part of the eigenenergies for a given set of 
quasimomenta $\{k_j\}$ is given by
\be \label{5.13a} 
P = \sum_{\{k_j\}} k_j,
\ee
\be \label{5.13b}
\Re(E(\{k_j\}) = \sum_{\{k_j\}} \Lambda(k_j).
\ee
The ground-state energy is formed by the combination of the quasienergies with 
negative values of $\Lambda(k)$, i. e.,
\be \label{5.14}
\Re (E_0) = \sum_{k\in \{\Lambda(k)<0\}} \Lambda(k).
\ee

We see directly from \rf{5.13} that in the simplest case $N=2$,  
$\Lambda(\lambda) = \frac{\lambda^2+1}{2}\Lambda(1)$, and from \rf{5.11}  the 
symmetry \rf{4.0} that happens in the OBC is not present in the case of PBC if $\lambda \ne 1$. This means that for $\lambda \ne 1$ the ground-state energy 
per site, has an anomalous behavior, being  distinct for  
different boundary conditions. 

Let us consider separately the quantum chains with distinct values of $N$.

{\it a)} $N=3$. In Fig.~6 we show the dispersion relations $\Lambda(k)$ for the cases $\lambda=2$, $\lambda=1$ and $\lambda=1/2$. The Fermi points 
($\Lambda(k_F)=0$) are given by
\be \label{5.15}
k_F = \arccos \left( \frac{-1\pm \sqrt{1+8\lambda^6}}{4\lambda^3} \right).
\ee
%%%%%%%%%%%%%%%%%
\begin{figure} [ht]
\centering
\includegraphics[width=0.45\textwidth]{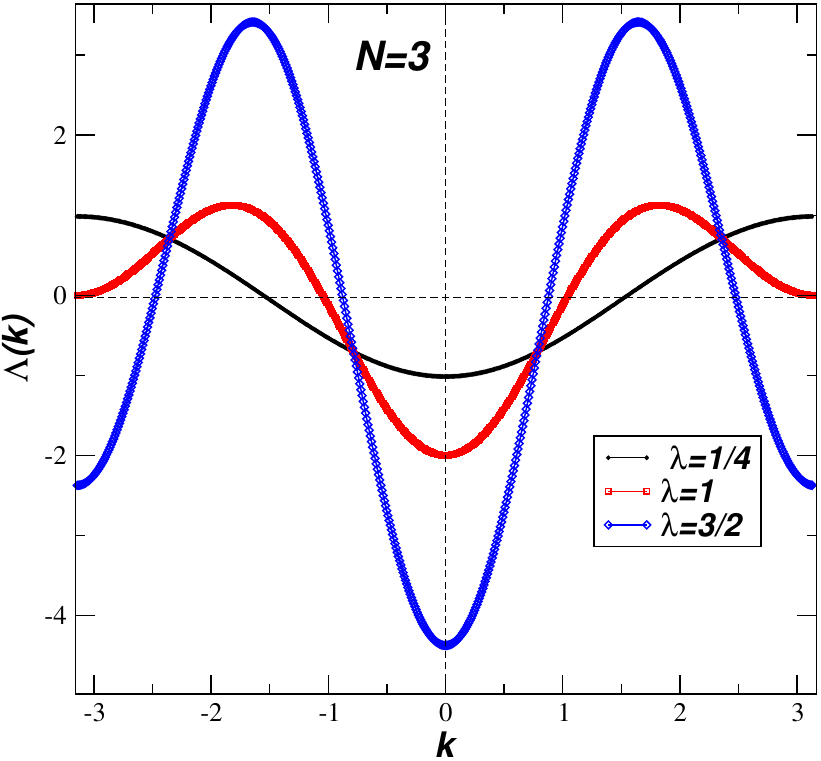}
\caption{  
Dispersion relation $\Lambda(k)$, given in \rf{5.13}, for the multispin $XX$ quantum chain with $N=3$, 
and some values of $\lambda$. The Fermi points are the ones where $\Lambda(k_F)=0$.}\label{fig6}
\end{figure}
%%%%%%%%%%%%%%%%%
We can show that for $\lambda \leq 1$ (see Fig.~6) we have two Fermi points ($k_F^{(1)},k_F^{(2)}$) while for $\lambda >1$ we have four of them 
($k_F^{(1)},\ldots,k_F^{(4)}$).  The ground-state energy is real and obtained 
from the addition of the quasienergies with $\Lambda(k_j) <0$, i. e,
\be \label{5.16}
E_0= - \sum_{\ell=1}^{N_{FP}/2} \sum_{k_F^{(2\ell-1)} \leq k_j \leq k_F^{(2\ell)}} 
(\cos k_j + \lambda^3 \cos (2k_j)),
\ee
where the number of Fermi points $N_{FP} =2$ if $\lambda \leq 1$ and 4 if 
$\lambda >1$. Since $\Delta k = k_{j+1} -k_j= \frac{2\pi}{L}$ we have in the 
bulk limit
\bea \label{5.17}
&&e_{\infty} (\lambda) = \lim_{L\to \infty} \frac{E_0}{L} \nonumber \\
&&= 
-\frac{1}{2\pi} 
\sum_{\ell=1}^{N_{FP}/2} 
\int_{-k_F^{(2\ell-1)}}^{k_F^{(2\ell)}} (\cos k +\lambda^3 \cos(2k))dk \nonumber \\
&& =-\frac{1}{2\pi} \sum_{\ell=1}^{N_{FP}} (-)^{\ell} (\sin k_F^{(\ell)} 
+ \lambda^3 \sin(2k_F^{(\ell)})).
\eea
At $\lambda=1$, $k_F^{(1)} = -\frac{\pi}{3}$, $k_F^{(2)}=\frac{\pi}{3}$ and 
\be \label{5.17a}
e_{\infty} = -\frac{1}{2\pi} \int_{-\frac{\pi}{3}}^{\frac{\pi}{3}} 
(\cos k + \lambda^3 \cos (2k) ) dk = -\frac{3\sqrt{3}}{4\pi},
\ee
that coincides with the conjectured value obtained in the OBC case. 
For $\lambda \neq 1$ the values obtained from \rf{5.16} and \rf{5.17} are
distinct from the prediction $-\lambda\frac{3\sqrt{3}}{4\pi}$ of the OBC 
(see \rf{4.0}), similarly as happens with the $Z(3)$ free-parafermionic 
Baxter model \cite{AB2}.

In order to verify the conformal invariance of the model, let us compute the 
conformal towers that should appear in the leading $L\to\infty$ finite-size 
behavior of the eigenenergies of the Hamiltonian.

Conformally invariant critical systems, with PBC, should have a ground-state 
energy $E_0^{(0)} (L)$, with the asymptotic finite-size behavior 
\cite{blote1,affleck1}
\be \label{5.18}
\frac{E_0^{(0)}(L)}{L} = e_{\infty} - \frac{\pi v_s c}{6L^2} + o(L^{-2}),
\ee
where $e_{\infty}$ is the ground-state energy per site in the bulk limit, 
$c$ is the conformal anomaly and $v_s$ the sound velocity, obtained from 
the energy-momentum dispersion relation. Moreover, for each operator 
\cite{cardy1986b} $\mathcal{O}_{\alpha}$ with dimension $x_{\alpha}$  in the 
operator algebra of the underlying conformal field theory, there exists an 
infinite tower of eigenstates in the quantum chain, that for $L$ sites 
and PBC should behave as:
\be \label{5.19}
E_{j,j'}^{\alpha} (L) = E_{0}^{(0)} + \frac{2\pi v_s}{L} 
(x_{\alpha} + j + j') + o(L^{-1}),
\ee

Let us consider initially the case $\lambda=1$. We take the finite-size 
sequences of even lattice sizes: $L=2\ell, \ell \in \mathbb{Z}$. The Fermi 
momentum are $k_F= \pm \frac{\pi}{3}$ and the ground-state energy is obtained 
by taking the symmetric distribution of $N_F = \frac{L}{3}$ fermions (even) 
and quasimomenta $k_j = \frac{2\pi}{L}(j+\frac{1}{2})$, $j=-\frac{L}{6},\ldots,
\frac{L}{6}-1$. This will give us from \rf{5.13a} a  zero momentum state 
($P=0$), and the sums in \rf{5.13b} give us the exact result:
\be \label{5.20} 
\frac{E_0^{(0)}} {L} = -\frac{\sqrt{3}}{4\sin(\pi/L)} \left(2 +\frac{1}{\cos(\pi/L)}\right).
\ee
The expansion for $L\to\infty$ give us
\be \label{5.21}
\frac{E_0^{(0)}}{L} = -\frac{3\sqrt{3}}{4\pi} - \frac{\sqrt{3} \pi}{4L^2} -
\frac{7\sqrt{3\pi^3}}{80L^4} + O(L^{-6}).
\ee
Comparing this result with the prediction \rf{5.18} we obtain 
$v_s c= \frac{3\sqrt{3}}{2}$.  On the other hand, the sound velocity can be 
obtained from the energy-momentum dispersion at the Fermi momentum
\bea \label{5.22}
v_s &=&\left. \frac{\partial E_0^{(0)}} {\partial k} \right|_{k=k_F} 
= 
\left. \frac{d\Lambda(k)}{dk} \right|_{k=k_F} \nonumber \\
&=& \sin{k_F} + 2\lambda^3 \sin(2k_F),
\eea
that for $\lambda=1$ give us $v_s = 3\sqrt{3}/2$.

We see from \rf{5.21} and \rf{4.44a} that  the ground-state energy per site 
is the same for the PBC and OBC at $\lambda=1$, differently  form the case 
the $Z(3)$ $p=1$ free-parafermionic Baxter chain, where they depend on the 
particular boundary condition \cite{AB2}. However even for $\lambda=1$ 
the sound velocity $v_s= 3\sqrt{3}/2$ for the PBC is half of the 
value obtained for the OBC \rf{4.15}. This anomalous behavior, even at 
$\lambda=1$, for the distinct boundaries is a consequence of the 
non-Hermiticity of the Hamiltonian. 

The eigensector containing the ground state have $N_F = \frac{L}{3}$ fermions 
and magnetization $m_z= -L/3$. We label the $U(1)$ symmetry charges relative 
to the ground state as 
\be \label{5.23}
Q = N_F - \frac{L}{3},
\ee
so that the ground state has zero momentum ad charge $Q=0$. In the 
ground-state sector ($Q=0$) we can create a state with momentum 
$P=\frac{2\pi}{L}$ (or $P=-\frac{2\pi}{L}$) by changing in the ground-state 
energy the quasimomentum $k_{\ell-1}$ (or $-k_{\ell}$) to the one with 
$k_{\ell}$ (or $k_{\ell+1}$), producing the mass gap
\bea \label{5.24}
\Re(G(\pm p)) &=& E_0^{(0)} - \Lambda(k_{\ell-1}) + \Lambda(k_{\ell }) 
\nonumber \\
&=&
E_0^{(0)} - \Lambda(-k_{\ell}) + \Lambda(-k_{\ell+1}),
\eea
whose $L$-large expansion give us
\be \label{5.25}
\Re(G(\pm p))= \frac{2\pi v_s}{L} - \frac{3\sqrt{3} \pi^3}{2 L^2} + O(L^{-4}).
\ee
Excited states with other momentum values, in the ground-state sector, are 
obtained by changing quasiparticles from bellow to above the Fermi 
momentum.

The lowest eigenenergy with $N_F=\frac{L}{3} +1$ ($N_F = \frac{L}{3}-1$) 
belonging to the sector $Q=1$ ($Q=-1$) is obtained by taking in \rf{5.13a} 
$k_j = \frac{2\pi}{L}$, $j=-\frac{L}{6},\ldots,\frac{L}{6} -1, \frac{L}{6}$ 
($j= -\frac{L}{6},\ldots,\frac{L}{6}-1$) and is given by

\be \label{5.26}
\frac{E_0^{(\pm 1)}}{L} = \frac{\sqrt{3}}{4L} 
\left(\tan(\frac{\pi}{L}) -
3/\tan(\frac{\pi}{L}) \right),
\ee
with the large-$L$ dependence
\be \label{5.27}
\frac{E_0^{(\pm 1)}}{L} = -\frac{3\sqrt{3}}{4\pi} +\frac{\sqrt{3}\pi}{2L^2} 
+ \frac{\sqrt{3}\pi^3}{10L^4} + O(L^{-6}),
\ee
giving us the gap
\be \label{5.28} \nonumber
E_0^{(\pm1)} - E_0^{(0)} = \frac{3\sqrt{3}\pi}{4L} + O(L^{-2})=\frac{2\pi v_s}{L} (\frac{1}{4}) + O(L^{-2}).
\ee

The prediction \rf{5.19} indicates the existence of a conformal operator with 
dimension $x=\frac{1}{4}$. The descendants of the operator will be related 
to the eigenenergies obtained by exciting the quasienergies that produce 
the lowest energy $E_0^{(\pm 1)}$.
The lowest eigenenergies with $U(1)$ charge $Q$ (or $-Q$), will have 
a zero momenta and are obtained by inserting (neglecting) symmetrically the 
quasi-particles forming the lowest energy configuration $E_0^{(0)}$ or 
$E_0^{(1)}$, depending if $Q$ is even or odd. A simple calculation give us 
the mass gap
\be \label{5.29}
E_0^{(Q)} -E_0^{(0)} = \frac{2\pi v_s}{L} \frac {Q^2}{4} + O(L^{-2}),
\ee
and from \rf{5.19} the related conformal dimensions are $Q^2/4$.

For a given sector $Q$, with a certain distribution $\{k_j\}$, the 
excitation where we take $\beta$ quasienergies near the Fermi momentum from 
the positive (negative) branch and insert them in the negative (positive) 
branch will give a set of mass gaps
\be \label{5.30}
\Re(E_0^{(Q,\beta)}) - E_0^{(0)} = \frac{2\pi}{L} v_s (\frac{Q^2}{4} + \beta^2),
\ee
giving us the conformal dimension $Q^2/4 + \beta^2$. The descendants of these 
dimensions are obtained by exciting the particles in the positive and negative 
branches.

These results imply that the model is described by a Gaussian conformal field 
theory with wave number $Q$ and vorticity $\beta$ \cite{kadanoff1,kadanoff2}, and 
dimensions given by
\be \label{5.31}
x_{Q,\beta} = \frac{2\pi}{L} v_s (Q^2 x_p + \frac{\beta^2}{4x_p}),
\ee
where $x_p=1/4$ and $Q,\beta=0,\pm1,\pm2,\ldots$. This is precisely the 
same operator content of the standard $N=2$ $XX$ model ($\lambda=1$) 
\cite{ope-cont1,ope-cont2}. The only differences are the non universal quantities 
$e_{\infty}$, $v_s$ and the $z$-magnetization associated to the $Q=0$ sector, 
that is zero in the standard $XX$ model and $L/3$ in the model with 
$N=3$ ($\lambda=1$).

For $\lambda <1$ we obtain a similar operator content an is the case 
$\lambda=1$, but with a sound velocity $v_s(\lambda)$ that depends on the 
Fermi momentum $\Lambda(k_F)=0$, and given by \rf{5.22}. For example for
\be \label{5.32}
\lambda^3 = \frac{\sqrt{5}-1}{1+\sqrt{5}} \approx 0.618034,\quad k_F = 
\frac{\pi}{5},
\ee
the ground state belongs to the sector of magnetization 
$m_z = m_z^{(0)}= -3L/5$, with the leading finite-size behavior
\be \label{5.33}
\frac{E_0}{L} = e_{\infty} -\frac{\pi v_s}{6L^2} + O(L^{-4}),
\ee
where 
\bea \label{5.34}
&&e_{\infty} =-\frac{2\sqrt{5+\sqrt{5}} + \lambda^3\sqrt{5-\sqrt{5}}} 
{4\sqrt{2}\pi}, \nonumber \\
&&v_s= 
\frac{\sqrt{5+\sqrt{5}} + 2\lambda^3\sqrt{5-\sqrt{5}}}
{4\sqrt{2}}.
\eea
The first excited state in the sector of magnetization $m_z = m_z^{(0)} \pm 1$,  similarly as in \rf{5.24}-\rf{5.25} give us the energy gap
\be \label{5.35}
E^{(\pm 1)} - E_0 = \frac{2\pi v_s}{L} \frac{1}{4}.
\ee

The results \rf{5.33} and \rf{5.35} indicate we have for $\lambda < 1$ the same conformal towers that appeared in the $\lambda=1$ case, only differing in the 
sound velocity $v_s(\lambda)$.

For $\lambda>1$ we have now 4 Fermi points (see Fig.~6). Our results indicate 
that we have a composition of two central charge $c=1$ theories. The sound 
velocities $v_s^{(1)}$ and $v_s^{(2)}$, in the branches 
$k_F^{(1)}$ and $k_F^{(4)}$ are distinct from the ones $v_s^{(2)}$ in 
$k_F^{(2)}$ and $k_F^{(3)}$. The ground-sate energy has the leading 
finite-size behavior
\be \label{5.36}
\frac{E_0}{L} = e_{\infty} - \frac{\pi}{6L^2} (v_s^{(1)} + v_s^{(2)}) + 
O(L^{-4}).
\ee
The excitations that will give the dimensions $x_p$ will give the same value 
$x_p=\frac{1}{4}$ in all branches. For example for $\lambda=(\sin(3\pi/14)/\sin(\pi/14))^{1/3}$, $k_F^{(2)}=-k_F^{(2)}= -2\pi/7$, and $k_F^{(1)}=-k_F^{(4)} \approx -0.7961934 \pi$. The contribution to the ground-energy for the branch 
$\frac{2\pi}{7} <k<\frac{2\pi}{7}$ has the leading behavior
\be \label{5.37}
\frac{E_0^{(1)}}{L} = \cos(\frac{\pi}{14}) 
\frac{\lambda^3 -2 +4\cos(\frac{\pi}{7})}{2\pi}
  - \frac{\pi v_s^{(1)}} {6L^2} + O(L^{-4}),\nonumber
\ee
with
\be \label{5.38}
v_s^{(1)} = \cos(\frac{3\pi}{14}) + 2\lambda^3 \cos(\frac{\pi}{14}) \approx 4.74437, \nonumber 
\ee
while the contribution from $-\pi \leq k \leq k_F^{(1)}$ and 
$\pi-k_F^{(4)} \leq k <\pi$ give us
\be \label{5.39}
\frac{E_0^{(2)}} {L} \approx  -0.683626 - \frac{\pi v_s^{(2)}} {6L^2} + 
O(L^{-4}), \nonumber
\ee
with 
\be \label{5.44}
v_s^{(2)} \approx 6.24521 . \nonumber
\ee

The excitations that contribute to the dimension $x_p$ give us $2\pi v_s^{(i)}/4$ ($i=1,2$), implying $x_p=1/4$.

These results imply that for $\lambda >1$ we have a mixture of two central charge $c=1$ theories with distinct sound velocities, giving us an effective theory with central charge $c=2$. Consequently we have at $\lambda=\lambda_c=1$ a phase 
transition from an effective theory with $c=1$ ($\lambda \leq 1$) to another one with $c=2$ ($\lambda>1$).

{\it {b)}} $N=4$. In Fig.~7 we show the dispersion relation $\Lambda(\lambda)$ 
for the cases $\lambda=0.25,1$ and $1.25$. From \rf{5.13} it follows  that for 
$\lambda <\lambda_c = \frac{1}{3^{1/4}} \approx 0.7598$ there exist only two 
Fermi points $k_F = \pm \frac{\pi}{2}$ in the Brillouin zone 
$-\pi \leq k < \pi$. The ground-state belongs to the sector with $m_z = L/2$ and have the leading finite-size behavior:
\bea \label{5.41}
&& \frac{E_0}{L} = e_{\infty} -\frac{\pi v_s}{6L^2} + O(L^{-4}), \nonumber \\
&& e_{\infty} = \frac{\lambda^4-3}{3\pi}; \quad v_s = 3\lambda^4-1.
\ee
%%%%%%%%%%%%%%%%%
\begin{figure} [ht]
\centering
\includegraphics[width=0.45\textwidth]{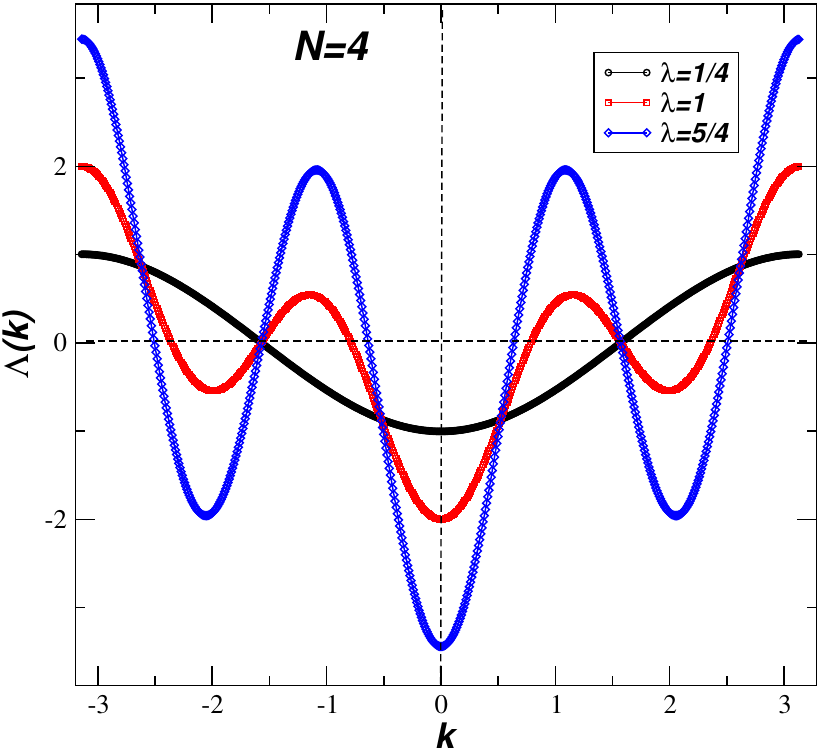}
\caption{  
Dispersion relation $\Lambda(k)$, given in \rf{5.13}, for the multispin $XX$ quantum chain with $N=4$, 
and some values of $\lambda$. The Fermi points are the ones where $\Lambda(k_F)=0$.}\label{fig7}.
\end{figure}
%%%%%%%%%%%%%%%%%
%parei
The first excited state in the sector with magnetization $m_z= L/2\pm1$ give us the dimension $x_p=1/4$, like in the former cases $N=2$ and $3$.

For $\lambda >\lambda_c$ appear four new Fermi points and an analysis similar 
as in the case $N=3$ show us that we have the finite-size leading behavior 
for the ground-state energy:
\be \label{5.43}
\frac{E_0}{L} = e_{\infty} -\frac{\pi}{6L^2} (\sum_{i=1}^6 v_s^{(i)}) 
+ O(L^{-2}),
\ee where 
\be \label{5.43}
v_s^{(i)} = |\sin k_F^{(i)} - 3\lambda^4 \sin(3k_F^{(i)})|,
\ee
 and  $k_F^{(i)}$ ($i=1,\ldots,4$) are the Fermi momenta ($\Lambda(k_F^{(i)})=0$). 

We then have a quantum chain ruled by an effective $c=3$ central charge 
theory, formed by the composition of 3 central charge $c=1$ theories, all 
of them with the polarization operator with dimension $x_p=1/4$.
 Actually these ground-states are indeed excited states of the standard $XX$ quantum chains, 
and the 
appearance of central charges proportional to the number of Fermi points (disjoint sectors of 
quasi-momenta) was observed in \cite{a1}, and also more recently in \cite{y1}. 

At $\lambda=1$ the Fermi points are $-3\pi/4,-\pi/2,\pi/2,3\pi/4$ and the 
value of $e_{\infty}=-\frac{2(1+2\sqrt{2})}{3\pi}$. This is distinct from the value $-\frac{2\sqrt{2}}{3\pi}$ given in \rf{4.44a} for the OBC. We see that 
 the anomalous behavior verifyed numerically for the $p=1$ $Z(3)$ 
free-parafermionic  Baxter model \cite{AB2} is observed analytically 
(even at $\lambda=1$) for the $N=4$-multispin $XX$ model. 

{\it c)} $N>4$. We conjecture that for small values of $\lambda <\lambda_c^{(1)}$ we 
have always a $c=1$ conformal spectrum and for large values $\lambda>\lambda_c^{(2)}$ the spectra if given by a mixture of ($N-2$)  $c=1$ theories, giving us  conformal towers 
of effective $c=N-1$ conformal theories. 
The dimension that generates all  the conformal dimensions (compactification 
ratio in the Coulomb gas language, or Luttinger parameter in spin liquid 
language) is always $x_p=1/4$, as in the standard $XX$ quantum chain.
 Actually, as we shall see in the next section, for the cases of the odd values of $N>3$, there exists intermediate 
phases with smaller central charges (see \rf{6.10a}-\rf{6.10b}).  Again, we verified that the energy per site $e_{\infty}$, even at $\lambda=1$ are distinct for the PBC and OBC cases.

\section{The entanglement entropies of the $XX$ multispin interaction 
quantum chains}

A direct test of the conformal invariance of a given critical quantum chain is 
the evaluation of the entanglement entropy obtained from the pure state 
density matrix
\be \label{6.1}
\rho= \ket{\Phi_L} \bra{\Phi_R},
\ee
where $\ket{\Phi_L}$ and $\bra{\Phi_R}$ are the left and right ground-state 
wave functions of the Hamiltonian. We split the chain of $L$ sites 
in two disjoint sublattices $A$ and $B$ containing $\ell$ and $L-\ell$ 
contiguous sites, respectively. The reduced density matrices of the 
subsystems $A$ and $B$ are obtained from the partial trace of the 
complementary subsystem, i. e.,  $\rho_A = \mbox{Tr}_B \rho$ and $\rho_B= \mbox{Tr}_A \rho$. The $\alpha$-R\'enyi entanglement entropy ($\alpha=1,2,\ldots$) of 
subsystem $A$ is defined as 
\be \label{6.2}
S_{\alpha} (\ell,L) = \frac{1}{1-\alpha} \ln [ \mbox{Tr }(\rho_A)^{\alpha }].
\ee

The limit $\alpha \to 1$ gives the von Neumann entanglement entropy 
\be \label{6.3} 
S_1(\ell,L) = - \mbox{Tr }(\rho_A \ln\rho_A).
\ee
The conformal invariant quantum chains, i.e., the ones ruled by an underlying 
conformal field theory, have a leading behavior as $L \to \infty$ ($\frac{\ell}{L}$ fixed) for the $\alpha$-R\'enyi entropy \cite{wilksec,calabrese-cardy1,calabrese-cardy2,korepin} 
\be \label{6.4}
S_{\alpha}(L,\ell) = \frac{c}{6\eta} (1 + \frac{1}{\alpha}) 
\ln \left[ \frac{\eta L}{\pi} \sin(\frac{\pi \ell}{L})\right] + a_{\eta}^{(\alpha)},
\ee
for $ \alpha = 1,2,\ldots$, $c$ is the central charge, $\eta=1$ ($\eta=2$) for 
PBC (OBC), and $a_{\eta}^{(\alpha)}$ is a non-universal constant. 

The models we are considering have a free-fermion eigenspectra. In this case 
there exist a standard method \cite{vidal1,peschel1,y2} to calculate the 
entropies $S_{\alpha}(L,\ell)$. The method is based on the evaluation 
of the eigenvalues $\nu_j$ ($j=1,\ldots,\ell$) of the correlation matrix 
$\mathbb{C}$, 
with elements 
\be \label{6.5}
C_{m,n} = \bra{\Phi_L} c_m^{\dagger} c_n \ket{\Phi_R}, 
\quad m,n=1,\ldots,\ell.  
\ee
The R\'enyi entanglement entropies are given by
\bea \label{6.6}
&&S_{\alpha}(L,\ell) = \frac{1}{1-\alpha} \sum_{j=1}^{\ell} \ln[
\nu_j^{\alpha}  + (1 -\nu_j)^{\alpha}], \nonumber \\
&&
\eea
and for the case $\alpha=1$ we have 
\bea \label{6.7}
&&S_1(L,\ell) =  \nonumber \\
&& -\sum_{j=1}^{\ell} \left[ \nu_j  \ln \nu_j  + (1-\nu_j) \ln (1-\nu_j)\right].
\eea

For simplicity we are going to present only the cases where the quantum chains 
are in a periodic lattice. In this case, from Sec. V, the left and right 
eigenvectors are given by

\be \label{6.8}  
\bra{\Phi_L} = \bra{0} \prod_{k \in \{k\}_0} \eta_k, \quad 
\ket{\Phi_R} = \ket{0} \prod_{k \in \{k\}_0} \eta_k^{\dagger}, 
\ee
where $\{\eta_k\}$ are the fermionic Fourier modes given in \rf{5.10a}. The 
set $\{k\}_0$ are formed by the quasimomenta in \rf{5.10c} 
defining the 
ground state, namely, the ones that give negative values for the quasienergies  $\Lambda(k)$, given in \rf{5.13}.

Inserting \rf{5.10c} in \rf{6.5} we obtain the elements of the correlation 
matrix
\be \label{6.9}
C_{m,n} = \frac{1}{4L} \sum_{k \in \{k\}_0} e^{-ik(m-n)},
\ee
the sets $\{k\}_0$  depend on the value of $N$ and the lattice 
size  
parity. 

The eigenvalues $\nu_j$ ($j=1,\ldots,\ell$) of the subsystem correlation 
matrices with elements $C_{m,n}^{(\ell)}$ ($m,n=1,\ldots,\ell$) give us the 
entanglement entropies $S_{\alpha}(L,\ell)$ in \rf{6.6} and \rf{6.7}.

In Fig.~8 we show the von Neumann entropy $S_1(L,\ell)$ as a function of $\ln[\frac{L}{\pi} \sin(\frac{\ell \pi}{L})]/3$ for the quantum chain with $N=3$ and 
$L=600$ sites, for some values of $\lambda$. The results for $S_2(L,\ell)$ and $S_3(L,\ell)$ for the values 
$\lambda=1/2$ and $\lambda=2$ are also shown in Fig.~9.
%%%%%%%%%%%%%%%%%
\begin{figure} [htb]
\centering
\includegraphics[width=0.45\textwidth]{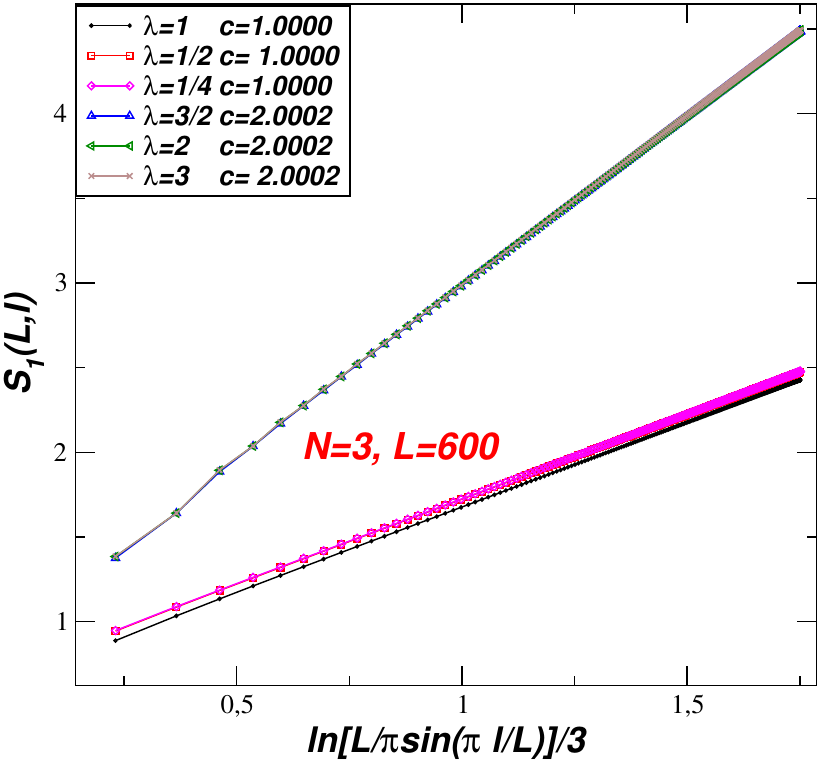}
\caption{  
The von Neumann entanglement entropy $S_1(L,\ell)$, as a function of $\ln[\frac{L}{\pi} \sin(\frac{\ell \pi}{L})]/3$, for some values of $\lambda$.
 The data are for the $XX$ multispin quantum chain with $N=3$, $L=600$ sites and PBC.} \label{fig8} 
\end{figure}
%%%%%%%%%%%%%%%%%
The estimated values of the central charge shown in these figures are obtained 
from the fit ($50 \leq \ell \leq 300$) with the expected form 
  \rf{6.4}
%parei aqui the 
quantum chains with $L=600$ sites. The data in Figs.~8 and 9 show a clear 
agreement with the prediction \rf{6.4} with $c=1$ for $\lambda \leq 1$ and $c=2$ for $\lambda>1$. This is even clear with the results of Fig.~10 where we show 
the estimates of the central charge $c$ as a function of $\lambda$. The values 
in this figure are obtained from the fit of $S_1(300,\ell)$ with \rf{6.4} 
by considering  $50\leq \ell \leq 150$. We clearly see a phase transition 
separating at $\lambda=\lambda_c=1$ the critical phases with $c=1$ and $c=2$, 
in agreement with the predictions of previous sections. In computing the 
entropies we should take into account that for $\lambda \leq \lambda_c$ we have 
only two Fermi points and for $\lambda >\lambda_c$ we have four of them. 

For general values of $N>3$ we also found a quite good agreement with the 
conformal invariance predictions \rf{6.4}.
%%%%%%%%%%%%%%%%%
\begin{figure} [htb]
\centering
\includegraphics[width=0.45\textwidth]{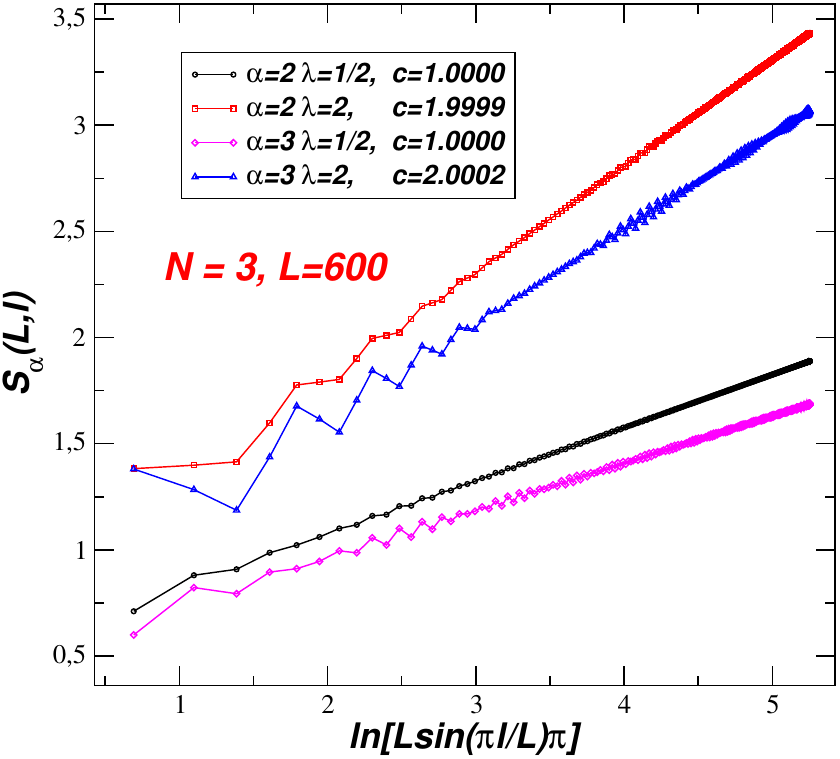}
\caption{  
The R\'enyi entanglement entropies  $S_2(L,\ell)$ and $S_3(L,\ell)$, as a function of $\ln(\frac{L}{\pi} \sin(\frac{\ell \pi}{L}))$.
 The data are for the $XX$ multispin quantum chain with $N=3$, $L=600$ sites and PBC.} \label{fig9} 
\end{figure}
%%%%%%%%%%%%%%%%%

In Fig.~11 we show our results for the central charge $c$, as a function of 
$\lambda$ for the model  with $N=4$ and $N=6$. For $N=4$ (open circles) 
the phase transition 
happens at $\lambda=\lambda_c=\frac{1}{3^{1/4}} \approx 0.7598$, 
separating a phase where $c=1$ from a phase where $c=N-1=3$. For $N=6$ 
(asteristiks) the phases are 
$c=1$ and $c=N-1=5$, and the transition parameter is 
$\lambda=\lambda_{c_1} \approx 0.9634$.

%%%%%%%%%%%%%%%%%
\begin{figure} [htb]
\centering
\includegraphics[width=0.45\textwidth]{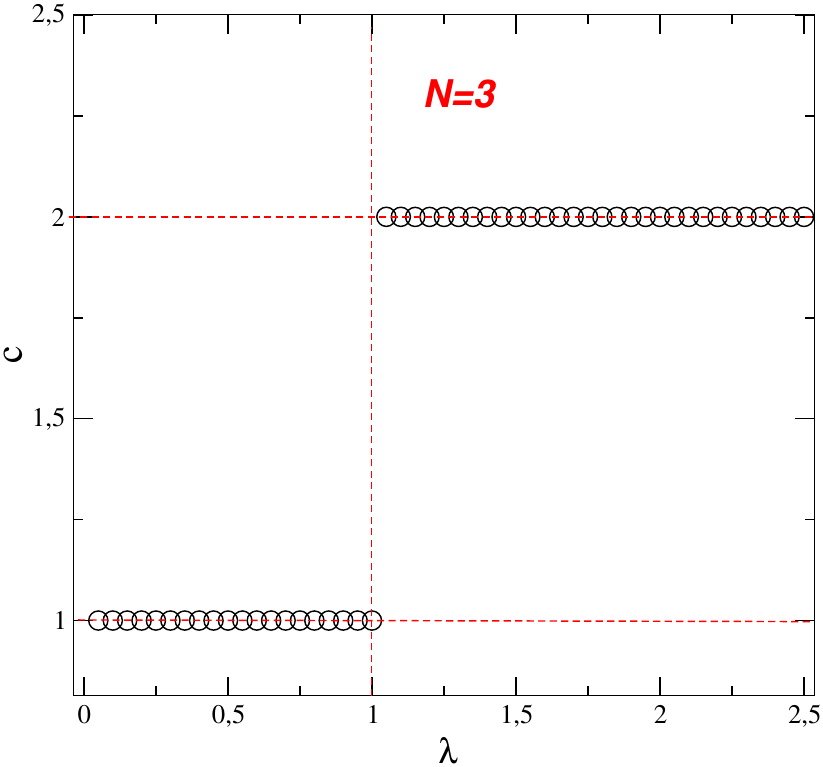}
\caption{  
Estimated values of the central charge $c$ as a function of $\lambda$ for the 
multispin $XX$ quantum chain with $N=3$ and PBC (see text).} \label{fig10}
\end{figure}
%%%%%%%%%%%%%%%%%
%%%%%%%%%%%%%%%%%
\begin{figure} [htb]
\centering
\includegraphics[width=0.45\textwidth]{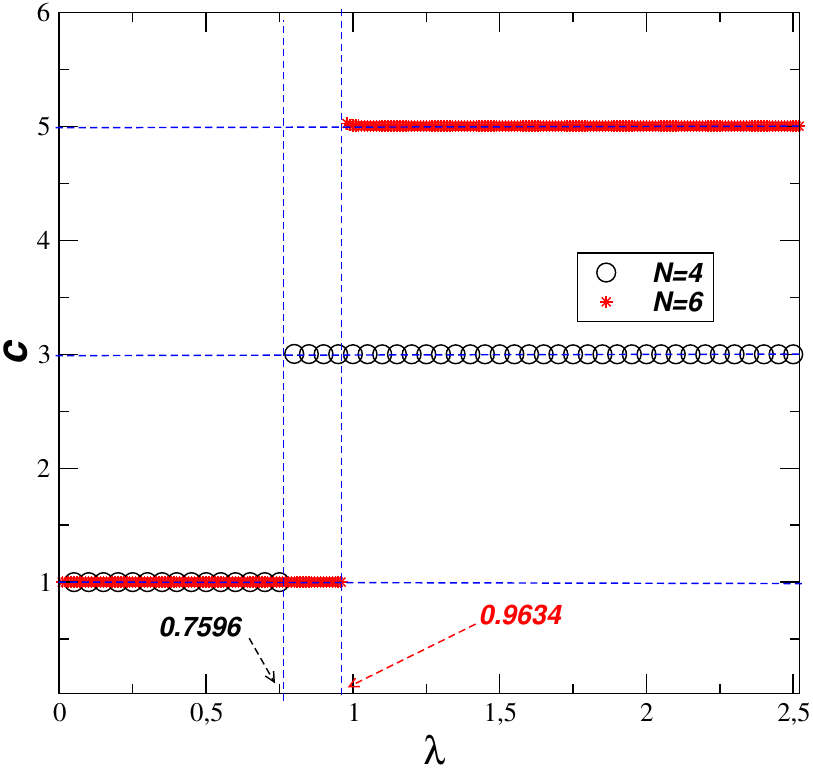}
\caption{  
Estimated values of the central charge $c$ as a function of $\lambda$ for the 
multispin $XX$ quantum chains with $N=3$ and $N=6$, with PBC (see text).} \label{fig11}
\end{figure}
%%%%%%%%%%%%%%%%%
%%%%%%%%%%%%%%%%%
\begin{figure} [htb]
\centering
\includegraphics[width=0.45\textwidth]{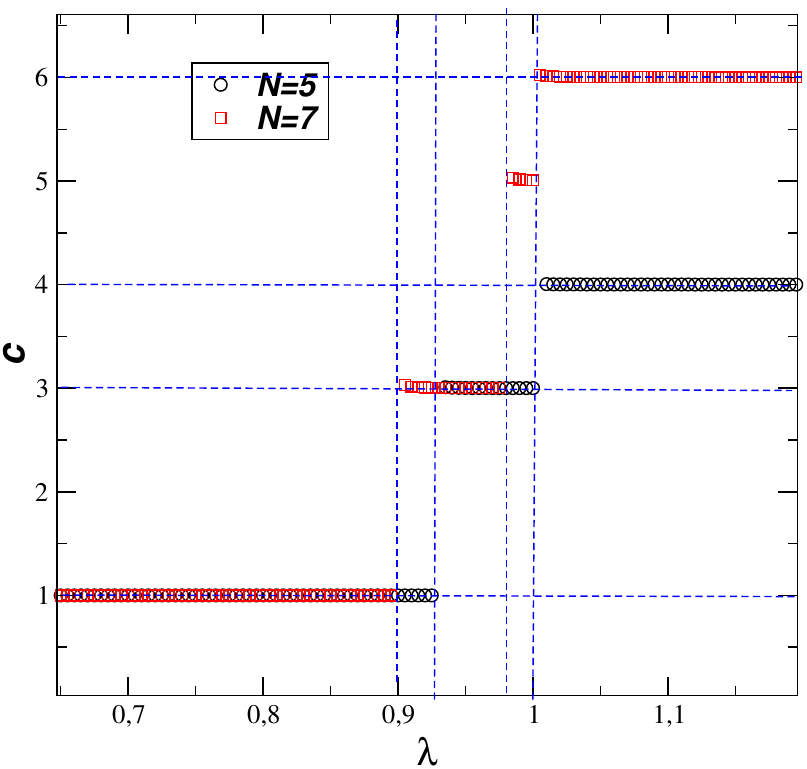}
\caption{  
Estimated values of the central charge $c$ as a function of $\lambda$ for the 
multispin $XX$ quantum chains with $N=5$ and $N=7$ with PBC (see text).} \label{fig12}
\end{figure}
%%%%%%%%%%%%%%%%%

In fig.~12 we show the central charge estimates for $N=5$ and $N=7$. 
We obtain the estimates from the fit of $S_1(900,\ell)$ ($50 <\ell<450$), with  the expression \rf{6.4}.
The model with $N=5$ (open circles) show three phases:  a phase for $0<\lambda\leq \lambda_{c_1}$ where the model has a central charge $c=1$, an intermediate phase for 
$\lambda_{c_1} <\lambda \leq \lambda_{c_2}$ where $c=3$ and a phase, for $\lambda >\lambda_{c_2}$ where $c=N-1=4$.  In the case $N=7$ (open squares)  the model has 4 phases: for 
$0<\lambda \leq \lambda_{c_1}$ the model has $c=1$, 
for $\lambda_{c_1} <\lambda \leq \lambda_{c_2}$ the phase has $c=3$, for 
$\lambda_{c_2} <\lambda\leq \lambda_{c_3}$ the phase has $c=5$ and for 
$\lambda>\lambda_{c_4}$ the phase has a central charge $c=N-1=6$. 
The 
phase transition points are
\be \label{6.10a}
 \lambda_{c_1} = 0.92645, \quad \lambda_{c_2}= 1,
\ee
for $N=5$, and for $N=7$ 
\be \label{6.10b}
 \lambda_{c_1} =0.89975 \quad \lambda_{c_2}=0.97899 \quad 
\lambda_{c_3} = 1.
\ee

Actually the phase transition points, separating conformal phases with distinct central charges, are precisely the ones where the number of Fermi 
points $N_{FP}$  changes in the dispersion relation. The  central charge is $c=N_{FP}/2$, in agreement with the results of Sec. V.

\section{Conclusions}

In this paper we study the spectral properties of two large families of free-particle quantum chains with 
multispin interactions. They are considered free because their energies are given by the sum of 
independent (free) pseudo-energies. In the first family we have parafermionic quantum chains 
with $Z(N)$ symmetry and ($p+1$)-multispin interactions ($p=1,2,\ldots$). The pseudo-energies 
forming the eigenenergies of the Hamiltonians, satisfy a $Z(N)$-circle exclusion contraint that 
generalizes the fermion exclusion principle for $Z(2)$. In the second family the models are 
$N$-multispin $XX$ models with a $U(1)$ symmetry, and described by a free-fermionic eigenspectrum.

The eigenspectra of both models with OBC are described in terms of the same pseudo-energies. 
These energies are exactly calculated from the roots of special polynomials. 
In their phase diagram  there exists a multicritical point with a dynamical 
critical exponent $z=(p+1)/N$. In the particular case where $N=p+1$ we have $z=1$, as in 
conformally invariant quantum chains. Our studies, when both models are in the  OBC
geometry, indicate that at those multicritical points the quantum chains are 
conformally invariant. The conformal invariance was tested by exploiting its consequences 
to the leading finite-size properties of the quantum chains in the finite geometry. 
These tests were done either analytically of with high numerical precision. The 
pseudo-energies for the OBC case are obtained from the roots of special polynomials with 
a known recursion relation.  We use a powerfull method \cite{powerfull1} that allow us 
to calculate the low-lying energies up to lattice sizes $\sim 10^9$. The numerical tests was 
done for the $Z(N)$ and $XX$ models with several values of $N$.  

For arbitrary $N$ our results indicate that indeed the models are described, at the 
multicritical points by a   conformal field theory. The sound velocity  and 
some of 
   the surface exponents were evaluated exactly.

For the case of periodic lattice (PBC) the situation is distinct. Due to the 
non-Hermiticity of 
the models we have quite distinct physics in the different geometries. For the isotropic 
models ($\lambda_1=\lambda_2=\ldots=\lambda$) the parameter $\lambda$ is just a harmless 
overall scaling factor for the models with OBC. However in the PBC case by changing $\lambda$ 
the models, although being critical as in the OBC, undergoes phase transitions. 

It is difficult to calculate the eigenspectra of the $Z(N)$ parafermionic quantum chains 
with PBC. This is not the case for the related $N$-multispin interacting $XX$ models, since 
in this case, thanks to  the standard Jordan-Wigner transformation, the Hamiltonian 
is a sum of bilinear fermion operators. In the case where the chain is traslational 
invariant, as happens for the isotropic model with PBC, the diagonalization follows from a 
Fourier transform, and all eigenfunctions are given by the composition of  Fourier 
modes. Actually this imply  a quite general result:

{\it All the translational invariant Hamiltonians (Hermitian or not) that can be expressed 
after a Jordan-Wigner transformation, in  a bilinear form will share the same 
eigenfunctions (not the eigenvalues), and consequently they commute among themselves.} 
The commutation follows directly from the fact that the general Hamiltonian has the form:
\beq \label{7.1}
H_L = \sum_{\ell=1}^L A_{\ell} h_{\ell}, \quad h_{\ell} = 
\sum_{\i=1}^L c_i^{\dagger} c_{\ell +1},
\eeq
with $\{A_{\ell}\} \in \mathbb{C}$, $\ell=1,\ldots,L$,  $c_{i+L} = c_i$, 
and $[h_{\ell},h_m]=0$.

This means that the study of all the wave functions of a simple model, 
like the standard two-body $XX$ model is equivalent to the study of 
all the eigenfunctions of the general free-fermion quantum chain \rf{7.1}.

The $N$-multispin $XX$ models with PBC and isotropic coupling $\lambda$, considered 
in this paper, are particular cases of \rf{7.1}. Our results of section 5 and 6 
indicate that for the periodic lattices the models undergo phase transitions as 
we change the value of $\lambda$. The finite-size behavior of the eigenspectra, 
for the models with $N=3,4,5$ and $6$, indicate that in general the models 
are critical and conformally invariant. In each phase the models have distinct 
central charges, whose values depend on the value of $\lambda$ and $N$. These 
phases appear because the models, although non-Hermitian, are described by 
Fermi surfaces and the number of Fermi points $N_{FP}$ depend on the particular value 
of $\lambda$ for a given $N$-multispin $XX$ quantum chain. The central charge 
has the value $c=N_{FP}/2$. It is important to mention that the energy 
per site  $e_{\infty}$ of the homogeneous models are the same for the periodic 
and open boundary cases, only when $N=2$ where the Hamiltonian is Hermitian. For 
$N\ge 3$ they show distinct values for the different boundary conditions, similarly as happens for the $Z(N)$ free-parafermion Baxter quantum chains, for 
$N\ge 3$ \cite{AB2}.

As a general scenario our results indicate that for small values of 
$\lambda<<1$ the models are always in a phase with the central charge $c=1$, and for $\lambda>>1$ the models are 
in a phase with central charge $c=N-1$. For general values of $N$ 
the  models show intermediate phases with integer values of the central 
charge ($1<c<N-1$), that are formed by independent  compositions of 
$c=1$ theories, all of them with the lowest conformal dimension $x_p=1/4$ 
(see Sec.~5 and 6). 

We conclude by stressing that all these ground states with distinct values 
of the central charge are also excited states of the general models 
\rf{7.1} with $\{A_{\ell}\}$ arbitray. An interesting question for the 
future concerns the phase diagram of the $Z(N)$ $N$-multispin models 
with PBC. Are these multiple phases also present?

\begin{acknowledgments}
We thank
 discussions with Jos\'e A. Hoyos.
This work was supported  in part by the Brazilian agencies FAPESP (Proc.2015/23849-7), CNPq and CAPES. 
\end{acknowledgments}


\begin{thebibliography}{99}


\bibitem{lieb}
T.~Schultz, D.~Mattis and E.~H.~Lieb,
``Two-dimensional Ising model as a soluble problem of many fermions,''
{\em Rev. Mod. Phys. \textbf{36} (1964) 856}.

\bibitem{pfeuty} P. Pfeuty, ``The one-dimensional Ising model with a transverse field," {\em Ann. Phys. \textbf{57} (1970) 79}.

\bibitem{fendley2}
P.~Fendley, ``{Free fermions in disguise}'',
{{\em J. Phys.} {\bfseries
  A52} 33 (2019) 335002},
{\ttfamily arXiv:1901.08078}.

\bibitem{baxter1}
R.~J. Baxter, ``{A simple solvable $Z_N$ Hamiltonian},''
{{\em Phys. Lett.}
  {\bfseries A140} (1989) 155}.

\bibitem{baxter2}
R.J.~Baxter, ``{Superintegrable chiral Potts model: Thermodynamic properties,
  an inverse model and a simple associated Hamiltonian},''
{{\em J. Stat. Phys.} {\bfseries
  57} (1989) 1}.

\bibitem{fendley1}
P.~Fendley, ``{Free parafermions}'',
 {{\em J. Phys.}
  {\bfseries A47} no.~7, (2014) 075001},
{{\ttfamily arXiv:1310.6049
  [cond-mat.stat-mech]}}.

\bibitem{baxter3}
R.~J. Baxter, ``The $\tau_2$ model and parafermions,''
{{\em J. Phys.} {\bfseries A47} (2014)
  315001}, {\tt arXiv:1310.7074 [cond-mat.stat-mech]}.

\bibitem{perk1}
H.~Au-Yang and J.~H.~H. Perk, ``Parafermions in the $\tau_2$ model,''
{{\em J. Phys.} {\bfseries A47} (2014)
  315002}, {\tt arXiv:1402.0061}.

\bibitem{perk2}
H.~Au-Yang and J.~H.~H. Perk, ``Parafermions in the tau-2 model II,'' 2016,
{\tt arXiv:1606.06319 [math-ph]}.

\bibitem{AB1}
F.~C. Alcaraz, M.~T. Batchelor, and Z.~-Z. Liu, ``Energy spectrum and critical
  exponents of the free parafermion $Z_N$ spin chain,''
{{\em J. Phys.} {\bfseries A50} (2017) 16LT03},
{\tt arXiv:1612.02617 [cond-mat.stat-mech]}.

\bibitem{AB2}
F.C.~Alcaraz and M.T.~Batchelor, ``Anomalous bulk behavior in the free
  parafermion Z(N) spin chain,''
{{\em Phys. Rev. E} {\bfseries 97} (2018) 062118 }, {\tt arXiv:1802.04453 [cond-mat.stat-mech]}.

\bibitem{AP1}
F.C.~Alcaraz and R.A.~Pimenta, ``Free fermionic and parafermionic quantum
spin chains with multispin interactions
,'' {{\em Phys. Rev. B} {\bfseries 102} (2020) 121101(R)}, {\ttfamily arXiv:2005.14622 [cond-mat.stat-mech]}.

\bibitem{AP2}
F.C.~Alcaraz and R.A.~Pimenta, ``Integrable quantum spin chains with free fermionic and parafermionic spectrum,''{{\em Phys. Rev. B} {\bfseries 102} (2020) 
235170}, {\ttfamily arXiv:2010.01116 [cond-mat.stat-mech]}.


\bibitem{network} S.~J. Elman, A.~ Chapman, and S.~T.~ Flammia,''Free fermions behind the disguise'', {{\em Commun. Math. Phys.} {\bfseries 388}, 969} (2021)
, {\ttfamily arXiv:2012.07857  [quant-ph]}; A.~Chapmam, S.~T.~Elman, and A.~J.Koll\'ar,
``Free-Fermion subsystem codes'', {{\em PRX Quantum} {\bfseries 3},030321} (2022); A.~Chapman, S.~J.~Elman, and
R.~L.~Mann,
``A unified graph-theoretic framework for Free-Fermion solvability'', {\ttfamily arXiv:2305.15625}.


%\bibitem{network} S.~J. Elman, A.~ Chapman, and S.~T.~ Flammia,''Free fermions behind the disguise'', {\ttfamily arXiv:2012.07857  [quant-ph]}.


\bibitem{circuits-pozsgay}
B. Pozsgay,``Quantum circuits with free fermions in disguise'',
{\ttfamily arXiv:2402.02984  [quant-ph]}.

\bibitem{fendley-pozsgay} 
P. Fendley and B. Pozsgay,
``Free fermions with no Jordan-Wigner transformation'',
{\ttfamily arXiv:2310.19897 [Cond-mat.stat-mech]}.

\bibitem{ising-analogues}
F.C.~Alcaraz, R.A.~Pimenta and J.~Sirker,
``Ising Analogs of Quantum Spin Chains with Multispin Interactions,``
 {{\em 
Phys. Rev. B} {\bfseries 107} (2023) 235136}, {\ttfamily arXiv:2303.15284 [cond-mat.stat-mech]}.

\bibitem{PT1} 
P.D.~Mannheim, '' PT symmetry as a necessary and sufficient condition for 
unitary time evolution'', {{\em Phil. Trans. R. Soc. A} (2023) 371:20120060}.

\bibitem{AP3}
F.C.~Alcaraz and R.A.~Pimenta, ``Free-parafermionic $Z(N)$ and 
free-fermionic $XY$ quantum chains'',{{\em Phys. Rev. E} {\bfseries 104} (2021) 054121}, {\ttfamily arXiv:2018.04372 [cond-mat.stat-mech]}.


\bibitem{powerfull1}
F.C.~Alcaraz, J.A,~Hoyos  and R.A.~Pimenta, ``Powerfull method to evaluate the gaps of free-particle quantum critical chains'',{{\em Phys. Rev. B} {\bfseries 104} (2021) 174206}, {\ttfamily arXiv:2109.01938 [cond-mat.stat-mech]}.

 \bibitem{randonp2}
F.C.~Alcaraz, J.A.~Hoyos  and R.A.~Pimenta'
``Random free-fermion quantum spin chain with multispin interactions'',
{{\em Phys. Rev. B} {\bfseries 108} (2023) 214413},
{\ttfamily arXiv:2308.16249 [cond-mat.dis-nn]}.

\bibitem{bat-pt} 
R.A.~Henry and M.T.~Batchelor,
``Exceptional Points in the Baxter-Fendley Free Parafermion Model'',
{{\em Sci Post} {\bfseries 15} (2023) 016}.

\bibitem{abb} F.~C.~Alcaraz, M.~N.~Barber, and  M.~T.~Batchelor,
 `` Conformal Invariance, the $XXZ$ Chain
and the Operator Content of Two-Dimensional Critical Systems,'' 
{\em  Ann. Phys. (N.Y.) \textbf{182} (1988) 280}.

\bibitem{cardy1986b} J.L~Cardy, ``Operator Content of two-dimensional 
conformally invariant theories,''{{\em Nucl. Phys. B} {\bfseries 270} (1986) 
186}.

\bibitem{laguerre} Q.I.~Rahman and G.~Schmeisser, {\em Analytic Theory of 
Polynomials}, edited by H. G. Dales and P. M. Neumanni, London Mathematical Society Monographs, (Clarendon Press, Oxford, 2002)Academic, New York, 1983), Vol. 26.

\bibitem{blote1} H.W.J.~Bl\"ote,J.L.~Cardy and M.P.~Nightingale, ``Conformal 
invariance, the central chargem and universal finite-size amplitudes at 
criticality,''{{\em Phys. Rev. Lett} {\bfseries 56} (1986) 742}.

\bibitem{affleck1}
I.~Affleck, ``Universal term in the free energy at a critical point and conformal anomaly,''{{\em Phys. Rev. Lett} {\bfseries 56} (1986)  746}.


\bibitem{tit}
L. Titvinidze and G. Japaridze,``Phase diagram of the spin extended model'',
{{\em Eur. Phys. J. B} {\bfseries 32} (2003) 383}.

\bibitem{kadanoff1}
L.P.~Kadanoff, 
``Multicritical Behavior at the Kosterlitz-Thouless Critical Point'',
{{\em Ann. Phys. (N.Y.)} {\bfseries 120} (1979) 39 }.

\bibitem{kadanoff2} L.P.~Kadanoff and A.~Brown,  ``Correlation functions on the critical lines of the Baxter and Ashkin-Teller models'',{{\em Ann. Phys. (N.Y.)} {\bfseries 121} (1979) 318 }.


\bibitem{ope-cont1} F.C.~Alcaraz, M.~Baake and V. Rittenberg, ``Operator Content of the $XXZ$ chain,''{{\em J. Phys. A} {\bfseries 21}(1988) L117}.

\bibitem{ope-cont2} F.C.~Alcaraz, U.~Grimm and V. Rittenberg, ``The $XXZ$  
Heisenberg Chain, Conformal Invariance and the Operator Content of $c<1$ 
Systems,''{{\em N. Phys. B } {\bfseries 316}(1989) 735}.

\bibitem{a1}
V.~Alba, M.~Fagotti, and P.~Calabrese. ``Entanglement entropy
of excited states'',{{\em J. Stat. Mech.}}, (2009) P10020, 2009,{\tt arxiv:0909.1999 [cond-mat.stat-mech]}.


\bibitem{y1}
Yi-Bin Guo {\it et al},``Entanglement entropy of non-Hermitian free fermions''
{{\em J. Phys. Cond. Mat.} {\bfseries 33} (2021) 475502}.

\bibitem{wilksec} C.~Holzhey, F.~Larsen, and F.~Wilczek, 
``Geometric and Renormalized Entropy in Conformal Field Theory,''{{\em N. Phys. B } {\bfseries 424}(1994) 443}.

\bibitem{calabrese-cardy1}
 P.~ Calabrese and J.L.~Cardy, 
``Entanglement Entropy and Quantum Field Theory,''{{\em J. Stat. Mech. } {\bfseries P06002}(1994) 443}, {\ttfamily arXiv:0405152 [hep-th]}.


\bibitem{calabrese-cardy2}
 P.~ Calabrese and J.L.~Cardy, 
``Entanglement Entropy and Conformal Field Theory,''{{\em J. Phys. A} {\bfseries 42}(2009) 504005}, {\ttfamily arXiv:0905.4013  [cond-mat.stat-mech]}.

\bibitem{korepin}
 V.~Korepin, 
``Universality of Entropy Scaling in One Dimensional Gapless,''{{\em Phys. Rev. Lett.} {\bfseries 92}(2004) 096402}, {\ttfamily arXiv:0311056  [cond-mat.str-el]}.


\bibitem{vidal1}
G.~Vidal, J.I.~Latorre, E.~Rico and A.~Kitaev ,
``Entanglement in Quantum Critical Phenomena,''{{\em Phys. Rev. Lett.} {\bfseries 90}(2003) 227902}, {\ttfamily arXiv:0211074  [quant-ph]}.

\bibitem{peschel1}
I.~Peschel,
``Calculation of reduced density matrices from correlation functions,''{{\em J. Phys. A } {\bfseries 36}(2003) L205}, {\ttfamily arXiv:0212631  [cond-mat]}.

\bibitem{y2}
P-Y. Changm J-S.You, X. Wen and S. Ryu, ``Entanglement spectrum and entropy in topological 
non-Hermitian systems and nonunitary conformal field theory'' {{\em Phys. Rev. Res,} {\bfseries 2} (2020) 033069}.

\end{thebibliography}
\end{document}